\documentclass[aps,prd,a4paper,twoside,notitlepage,floatfix,nofootinbib,preprintnumbers,eqsecnum]{revtex4-1}

\usepackage[paperwidth=210mm,paperheight=297mm,centering,hmargin=2cm,vmargin=2.5cm]{geometry}

\usepackage{pstricks}
\usepackage{calc}
\usepackage{graphicx,graphics}
\usepackage{subfigure}
\usepackage{amsmath,amssymb,latexsym}
\usepackage{slashed}

\newcommand{\p}{\partial}
\newcommand{\eq}{\begin{equation}}
\newcommand{\eqe}{\end{equation}}
\newcommand{\nn}{\nonumber}
\newcommand{\eqa}{\begin{eqnarray}}
\newcommand{\eqae}{\end{eqnarray}}
\newcommand{\nomin}{\hphantom{-}}

\newcommand{\Mp}{M_{\mathrm{pl}}}
\newcommand{\sech}{\mathrm{sech}}

\newcommand{\WidestEntry}{$lon_1$}%
\newcommand{\SetToWidest}[1]{\makebox[\widthof{\WidestEntry}]{$#1$}}%

\def\Xint#1{\mathchoice
{\XXint\displaystyle\textstyle{#1}} 
{\XXint\textstyle\scriptstyle{#1}} 
{\XXint\scriptstyle\scriptscriptstyle{#1}} 
{\XXint\scriptscriptstyle\scriptscriptstyle{#1}} 
\!\int}
\def\XXint#1#2#3{{\setbox0=\hbox{$#1{#2#3}{\int}$ }
\vcenter{\hbox{$#2#3$ }}\kern-.6\wd0}}
\def\ddashint{\Xint=}
\def\dashint{\Xint-}

\begin{document}

\author{Joel M. Weller\footnote{joel.weller@kit.edu}}
\address{Institute for Theoretical Physics, Karlsruhe Institute of Technology, 76128 Karlsruhe, Germany}

\title{Fermion condensate from torsion in the reheating era after inflation}

\preprint{KA-TP-16-2013}
\begin{abstract}

The inclusion of Dirac fermions in Einstein-Cartan gravity
leads to a 
four-fermion interaction mediated by non-propagating torsion,
which can allow for the formation of 
a Bardeen-Cooper-Schrieffer condensate.
By considering a simplified model in 2+1 spacetime
dimensions, we show that
even without an excess of fermions over antifermions,
the nonthermal distribution 
arising from preheating after inflation can give rise to a 
fermion condensate generated by torsion.
We derive the effective Lagrangian for the spacetime-dependent pair field describing the condensate
in the extreme cases of nonrelativistic and massless
fermions, and show that it
satisfies the Gross-Pitaevski equation for a gapless, propagating mode.
 
\end{abstract}

\maketitle

\section{Introduction}

Unlike scalar and vector fields, fermions do not sit nicely in the
standard metric formalism of general relativity.
Since spinors live in SU(2) one can instead introduce vierbeins\footnote{
We follow the sign convection used in \cite{ParkerToms}, 
corresponding to $---$ in the classification of Misner et al. \cite{MTW} and 
use uppercase latin indices for components of tensors in a local orthonormal frame. 
}
$e^I_\mu$,
which provide a connection between spacetime indices (Greek)
and the internal Lorentz indices (Latin). These are related to the metric by $g_{\mu\nu} = \eta_{IJ}e^I_\mu e^J_\nu$. 
The components of any
 tensor object can be expressed in an orthogonal basis or a coordinate basis. In the former case, the
 covariant derivative $\nabla_\mu$ is defined in terms of the spin connection $\omega_\mu{}^I{}_J$ so that
 \[
 \nabla_\mu X^I{}_J = \partial_\mu X^I{}_J +\omega_\mu{}^I{}_K X^K{}_J -  \omega_\mu{}^K{}_J X^I{}_K. 
 \]
The metric compatibility condition ($\nabla g = 0$) implies $\omega_{\mu IJ} = -\omega_{\mu JI}$.
For consistency in the case of a mixed basis (i.e. a combination of indices $I$ and $\mu$)
we require the tetrad postulate
\[
\nabla_\mu e_\nu^I = \partial_\mu e_\nu^I - \Gamma^\lambda_{\mu\nu}e_\lambda^I+\omega_\mu{}^I{}_{J}e^J_\nu = 0,
\]
to be satisfied, irrespective of the properties of the coordinate basis connection $\Gamma^\lambda_{\mu\nu}$. In particular, we can consider the situation in which the torsion tensor, defined by $T_{\mu\nu}{}^\lambda = \Gamma^\lambda_{\mu\nu} - \Gamma^\lambda_{\nu\mu}$, is nonzero.

Without fermions, it is the variation of the action with respect to the connection that gives the
torsionless condition. The covariant derivative of fermions necessarily involves the
spin connection, and thus when fermions are included there is an extra term,
which, upon repeating the procedure, gives rise to an
extra four-fermion interaction; an observation first noted by Kibble in 1960 \cite{Kibble:1961ba}.
This can be found by considering
both the standard Einstein-Hilbert action (written in terms of vierbeins) as well as the
Holst term, involving the dual of the Riemann tensor (cf. \cite{deBerredoPeixoto:2012xd,Magueijo:2012ug}, and references therein). The strength of this 
interaction depends on the coupling constant of this term, the Barbero-Immirzi parameter $\gamma$, which (if real) is unconstrained\footnote{
In four-dimensions, the fermion interaction induced in this way is A-A (A is the axial current)
It is also possible for the interaction to have V-A (vector-axial current) and V-V interactions, in the case of a
non-minimal coupling of the fermions to gravity. The interaction
strength is then also dependent on the non-minimal coupling constant.
}. 
However, the particular combination that is relevant for the
action is $G\gamma^2/(1+\gamma^2)$, so in any case it is extremely weak. For simplicity, in this work, the gravitational sector is described by the Einstein-Hilbert action. 

Given an extra interaction between fermions, one can investigate the consequences. There are two
approaches in the literature: either to consider the modifications to the fermionic fluid due to the self-interaction \cite{Dolan:2009ni,deBerredoPeixoto:2012xd,Khriplovich:2012yj,Khriplovich:2013tqa}
 or form a condensate \cite{Poplawski:2010jv,Poplawski:2011wj}.
The latter approach is of interest, especially in the case of neutrinos, as there is
the possibility of finding a connection between the neutrino mass scale and dark energy\footnote{
Current interest in neutrino condensates notwithstanding, study of the interaction between neutrinos 
and the gravitational field goes back at least to work by Brill and Wheeler in 1957 \cite{Brill:1957fx}.
}, thereby
at least easing the cosmological constant problem.
(See \cite{Caldi:1999db,Blasone:2004yh,Barenboim:2010db} for related work utilising other interactions.) 
With this aim,
Alexander and collaborators \cite{Alexander:2009uu,PhysRevD.81.069902,Alexander:2008vt,Alexander:2006we} have investigated the formation of a Bardeen-Cooper-Schrieffer (BCS) condensate.
They consider the
pairing of neutrinos in order to derive the gap equation that is related to the change in the vacuum energy of the
system. 

However, the analogy with condensed matter means that the neutrino BCS condensate model
requires the same basic conditions for the formation of a condensate as in low-temperature superconductors:
degenerate fermions, i.e. it is assumed that there is a substantial chemical potential.
In low-temperature superconductors,
a Fermi surface at $\epsilon = \mu$ can form in momentum space, below which the occupation number is approximately 1 and above which it is 0.
Cooper pairs form in a small region (of momentum space) in the vicinity of the Fermi surface where a phonon mediated interaction becomes positive. The
presence of an attractive force, no matter how small, leads to the formation of pairs.
In Alexander et al., the authors integrate out the fermions and write an effective potential for the pair field 
$\Delta$,
however, unlike the standard BCS case (where the interaction is attractive only in a finite region) this involves a divergent integral over momentum space.
Regularising the potential introduces a new scale that, although unconstrained, determines the
energy density of the condensate.
The motivation for treating the regularisation scale as physical is that the interaction is non-renormalisable (although this is not obvious from the original form of the action) but this results in a
loss of predictive power.\footnote{
Another problem is the violation of Lorentz invariance, which is inherent in the use of a homogeneous
chemical potential. So we must position ourselves in the cosmic frame, which is rather unsatisfactory. 
A related problem is that the condensate itself defines a preferred frame. It has been suggested by Alexander et al. that neutrino oscillations could
result from an MSW-like effect due to their interaction with the condensate. However, this would
require the torsion (i.e. a gravitational degree of freedom) to have flavour dependence.}

A serious obstacle is that the conditions for the formation of a BCS condensate via Cooper pairing do not appear to be satisfied in the early universe. The density is certainly high, so a gravitational strength interaction is more relevant that at the present time, however, the temperature is also extremely high, so $\mu\gg T$ would require an enormous excess of fermions. 
The chemical potentials of different species are tightly constrained: for charged species $\mu/T \sim n_p/n_\gamma \sim 10^{-10}$ and for neutrinos $|\mu_{\nu}/T| < 0.07$ \cite{Serpico:2005bc}.
While the universe expands adiabatically, the ratio $\mu/T$ is unchanged, so although the chemical potential is
larger in the early universe, it is not significant.

In search of a large chemical potential in the early universe, one could consider an
out-of-equilibrium situation in a yet earlier epoch. 
Models of inflation of all types face the problem of recovering the
initial conditions of the hot big bang model after the period of quasi-exponential expansion 
ceases. 
In scalar field models this can be realised with a period of reheating, in which the scalar field
(inflaton) decays perturbatively into lighter fields while undergoing oscillations about its potential.
It was realised by Kofman and collaborators that in addition,
the field can decay by parametric resonance, in which the number density of created particles
is amplified by an exponential factor in particular resonance bands \cite{Kofman:1997yn}. The decay products (which are necessary relativistic due to their lightness compared to the inflaton field) then subsequently thermalise.
At first it was thought that this process was only possible for bosons, as Pauli blocking places a bound on the occupation number of fermions. However, it was found \cite{Greene:2000ew}
not only that fermionic preheating is possible, but that it can be extremely efficient.
In the
case of an expanding universe, the created fermions can be thought of as
stochastically filling a Fermi sphere defined by $k_F = q^{1/4}m_\phi$, where $m_\phi$ is the
inflaton mass.
The parameter $q$ is defined by $q = h^2\phi_0^2/m_\phi^2$, where $\phi_0$
is the amplitude of the inflaton oscillations and $h$ the 
(Yukawa type) coupling between the inflaton and the fermions.

Models of far-from-equilibrium phenomena indicate that the particles
undergo prethermalisation on timescales dramatically shorter than the thermal equilibration time, giving rise to a constant `kinetic temperature' 
$T_{\rm kin} = (E_{\rm kin}(t)/E_{\rm kin, eq})T_{\rm eq}$, even though 
the system is not yet in thermal or chemical equilibrium \cite{Berges:2004ce}.
One can think of this as arising due to the loss of phase information resulting from
the smoothing of the rapidly oscillating fermion distribution function
generated by the inflaton.
In contrast, the `mode temperatures', defined by equating the 
mode numbers to a Fermi-Dirac distribution,
are not constant for all modes until thermalisation occurs,
so the distribution can differ significantly from a thermal one.
More detailed studies of fermions produced during preheating \cite{Berges:2010zv}
indeed show that
the distribution is approximately thermal for IR modes,
with a temperature determined by the total energy density of fermions.
However, for modes larger than the scale set by the amplitude of the background
oscillations $\phi_0$, the occupation falls off with a shape reminiscent of a Fermi-Dirac distribution
with a chemical potential.
Over time, the decay products of the inflaton will thermalise completely
and, except for changes due to, for example, baryogenesis,
 the final distribution will be determined almost exclusively by the
temperature.
Since the torsion-generated four-fermion interaction is not
confined to a small region in phase space, as in the BCS case, it is
interesting to consider the possibility that any Cooper pairs formed 
during such a nonthermal period could persist after thermalisation.

 In this paper we address the problem by considering a simplified model in 2+1 spacetime
dimensions.
Three dimensional gravity has been a focus of study for many decades due to its interesting and 
surprising properties (cf. \cite{Deser:1983tn,Carlip:2005zn} for discussions and further references).
Static and stationary solutions of Einstein-Cartan gravity in 2+1 dimensions have been investigated in \cite{Hortacsu:2008uy,Dereli:2010kv}.
In our case, the reduced dimensionality makes it possible to obtain analytical expressions for
the coefficients entering the effective action for the pair field.
We implement the effect of the nonthermal distribution by
 splitting the momentum integrals at the Fermi surface corresponding to 
 $k_F = q^{1/4}m_\phi$ and assigning high and low temperatures to the low and high momentum
 modes respectively. This produces an effect similar (but opposite) to 
$y$-distortion in the cosmic microwave background (CMB),
which occurs when Compton scattering becomes less efficient, leading to a decrease in 
temperature for low energy photons and a corresponding increase for high energy photons.

We begin in Sec. \ref{NonRelSec} with the simpler problem
of deriving the effective Lagrangian
for the spacetime dependent pair field in the extreme case of nonrelativistic fermions   
using this prescription (described in detail in Sec. \ref{NonRelEffectiveLagrangianDerivation}).
In Sec. \ref{RelSec} we
 derive the gravitational four-fermion interaction explicitly in 2+1 dimensions and comment on the possible interaction channels. 
Using this result, in Sec. \ref{MasslessCase} we
derive the effective Lagrangian for the opposite extreme --- massless fermions ---
and close in Sec. \ref{Discussion} with a discussion of the implication of the results.

\section{Non-relativistic fermions}\label{NonRelSec}

\subsection{Preliminaries}

As a stepping stone to understanding the effect of a nonthermal distribution of fermions
on the formation of a BCS condensate, we consider a nonrelativistic toy model
in 2+1 dimensions in which the temperature entering the momentum integrals differs for
large- and small-scale modes. Following the review paper by Schakel \cite{Schakel:1999pa}
we use the non-relativistic Lagrangian
\eq
\mathcal{L} = \psi_{\uparrow}^*[i\p_0 - \xi(-i\nabla)]\psi_\uparrow+
 \psi_{\downarrow}^*[i\p_0 - \xi(-i\nabla)]\psi_\downarrow +
 \lambda_0 \psi_{\uparrow}^* \psi_{\downarrow}^* \psi_{\downarrow} \psi_{\uparrow},
\eqe
where $\xi(-i\nabla)=\epsilon(-i\nabla)-\mu_0$ with $\epsilon(-i\nabla) = -\nabla^2/2m$
and
$\lambda_0$ is a positive coupling constant characterising the attractive
four-fermion interaction.
The metric signature is $(+,-,-)$.
Introducing the auxiliary field
\eq
\Delta=\lambda_0\psi_\downarrow\psi_\uparrow,
\eqe
and 
integrating out the fermions
one gets
\eq
Z = \int D\Delta^*D\Delta \exp\left( i S_{\rm eff}-\frac{i}{\lambda_0}\int_x |\Delta|^2 \right),
\eqe
where the effective action is given by the formal expression
\eq
S_{\rm eff} = -i{\rm tr} \int_x \int_k e^{ik\cdot x}\log\left( 
\begin{array}{cc} p_0 - \xi(\bf p) & \Delta \\ \Delta^* & p_0 + \xi(\vec p)  \end{array}
 \right)  e^{-ik\cdot x},
\eqe
and the integrals are abbreviated as $\int_x=\int_{t,\bf x}=\int dt d^2x$ and $\int_k=\int_{k_0,\bf k} = \int\frac{d^3k}{(2\pi)^3}$.
When the pair field is spacetime dependent we need to perform a derivative expansion on the
logarithm. Shifting the momentum operators to the right using
\eq \label{DerivativeExpansionDef}
f(x)p_\mu g(x) = (p_\mu - i\partial_\mu)f(x)g(x),
\eqe
where the derivative $\p_\mu = (\p_0,-{\bf \nabla})$ acts only on the next object to its right,
we obtain 
the following quadratic and quartic terms
\eqa
S_{\rm eff}^{(2)} &=& i \int_x \int_k
\frac{1}{k_0+\xi(\bf k)}\frac{1}{\tilde k_0-\xi(\bf \tilde k )}\Delta^*\Delta, 
\label{NonRel_S2}
\\
S_{\rm eff}^{(4)} &=& \frac{i}{2} \int_x \int_k \frac{1}{[k_0^2-\xi^2({\bf k})]^2}|\Delta|^4,
\eqae
where $\tilde  k_\mu = (k_0-i\p_0, {\bf k} +i{\bf \nabla})$.
The temperature is included by going to imaginary time $t\rightarrow -i\tau$
and substituting 
\[
\int \frac{dk_0}{2\pi} g(k_0) \rightarrow i\beta^{-1}\sum_n g(i\omega_n),
\]
where $n$ is an integer,
$\omega_n=\pi\beta^{-1}(2n+1)$ is the (fermionic) Matsubara frequency and $\beta=1/T$.
This gives the finite temperature (Euclidean) action
\eq\label{SEffE}
S_{\rm eff}^E = \int_0^\beta d\tau \int_{\bf x}\int_{\bf k} \beta^{-1}\sum_n
\left[ \frac{1}{i\omega_n+\xi(\bf k)}\frac{1}{i\omega_n+\p_\tau-\xi(\bf k +i\nabla)}\Delta^*\Delta 
+\frac{1}{2}\frac{1}{[\omega_n^2-\xi^2(\bf k)]^2}|\Delta|^4
 \right],
\eqe
where the Euclidean action is related to the Minkowski action by
\[
S= -i\int_0^\beta d\tau \int_{\bf x}  \mathcal{L}(-i\tau,{\bf x}) =    S^E.
\]

\begin{figure}
\subfigure{\includegraphics[width=7.9cm]{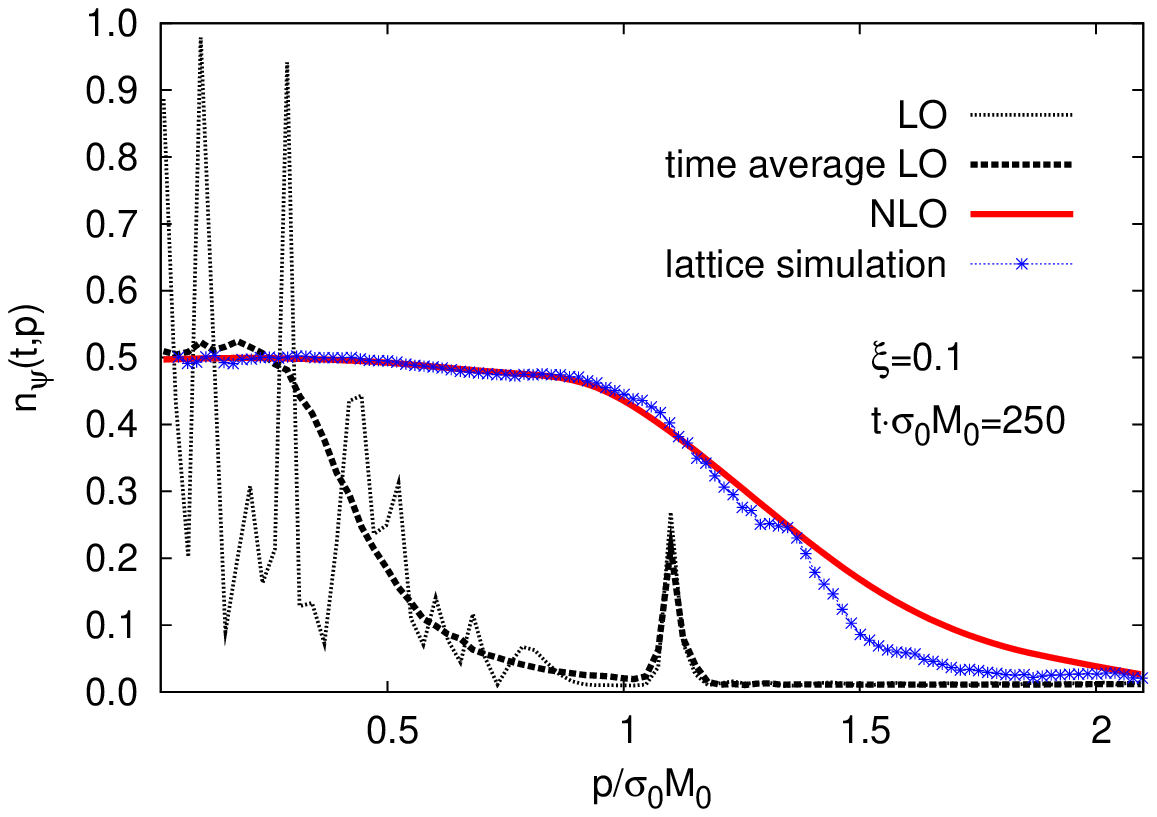}}
\subfigure{\raisebox{4.65mm}{\includegraphics[width=7.3cm,height=4.9cm]{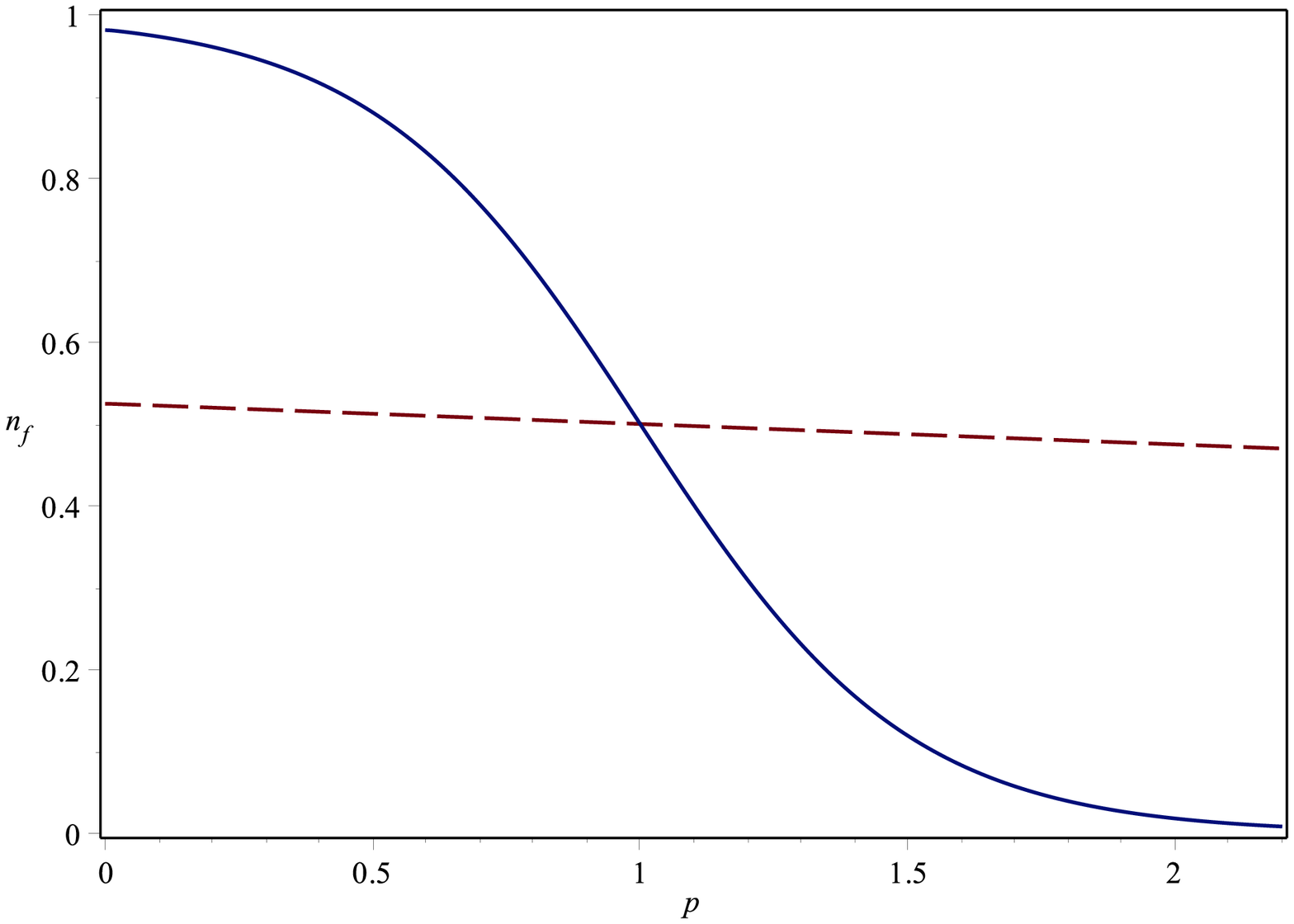}}}
 \caption{
Left: number distribution of massless fermions
produced after preheating (from Berges, Gelfand and Pruschke (2011) \cite{Berges:2010zv} with kind permission).
The different curves indicate the result of the
inclusion of leading order (LO) and next-to-leading order (NLO) effects in the
calculation.
Right: Distribution functions 
 $f(p)=[\exp(\beta (E(p)-\mu))+1]^{-1}$
(red, dashed) and
  $f(p)=[\exp(\alpha\beta (E(p)-\mu))+1]^{-1}$
(blue, solid)
with $\alpha=40$, $\beta=0.1$, $\mu=1$ and $m=0.01$.
} \label{f_plot}
\end{figure}

\subsection{Deriving the effective Lagrangian}
\label{NonRelEffectiveLagrangianDerivation}

Using the expressions for the sums over the Matsubara frequencies given
in the appendix we can expand the integrand to get a time-independent Lagrangian of the form
\eq\label{LagrangianAnsatz}
\mathcal{L}_{\rm eff} = \Delta^*(c\nabla^2+a)\Delta - \tfrac{1}{2}b|\Delta|^4.
\eqe

The distribution (see Fig. \ref{f_plot}) 
is such that low momentum modes approximately follow a 
thermal distribution with a large temperature related to the total energy, while high momentum states
are not yet filled. 
To take account of the distortion from a true thermal distribution, we split the momentum integral
into two parts and replace all instances of $\beta$ by
$\beta/\alpha$ for $k$ satisfying  $\xi = k^2/2m-\mu<0$
and $\alpha\beta$ for $k$ satisfying $\xi = k^2/2m-\mu>0$ i.e.
\eq
\beta(k) = \Bigg\lbrace \begin{array}{cc}
\beta/\alpha & \xi <0 \\
\alpha\beta & \xi >0
 \end{array}
 \label{TempPrescription}
\eqe 
The effect is similar (but opposite and more extreme) to $y$-distortion in the cosmic microwave background,
which occurs when Compton scattering becomes less efficient, leading to a decrease in 
temperature for low energy photons and a corresponding increase for high energy photons (cf. \cite{Chluba11012012}).
To model the sharp decrease in occupation number at $\xi=0$, we consider $\alpha\gg 1$. 
In particular, we assume 
\eq\label{c12Defs}
c_1 \equiv \frac{\mu\beta}{2\alpha} \ll 1, \qquad\qquad c_2 \equiv \frac{\mu\alpha\beta}{2} \gtrsim 1,
\eqe
which are used frequently in the following sections.
Thus, the effect of replacing the temperature
for high momentum modes by a smaller value is that the effect of the chemical potential
is no longer negligible.  
It should be stressed that the chemical potential term considered here
is unrelated to any excess of fermions, but is merely a parameter used to describe the
fermion distribution, and corresponds roughly to a value $\mu\sim k_F = q^{1/4}m_\phi$ below which the bulk
of the fermion modes were created during preheating.
Thermalisation in this situation would correspond to the limit $\alpha\rightarrow 1$.

Both $\mu$ and $\alpha$ have physical interpretations relating to the
underlying physics. In this simplistic scenario, the end-product of the non-perturbative
parametric resonance decay of the inflaton is represented by a two-parameter toy model.
$\mu$, which acts as the chemical potential, is related to the inflaton mass
because it is this mass scale that determines the limit in momentum space below which most of the
fermions are created by parametric resonance. $\alpha$ is not related directly
to the parameters in the inflaton Lagrangian, but instead depends on the physics of the decay 
process in a nontrivial way. 
It is a measure
of the extent to which parametric resonance prefers low-momentum fermions, and cannot be too large (or too small) without failing to accurately represent the phenomenology of the full calculation.

So if $\alpha$ is not undetermined, what value should it take?
Because $\alpha = 1$ corresponds to a thermal distribution, any 
deviation requires that $\alpha >1$; because the distortion has been
shown to be large in detailed lattice calculations, $\alpha$ must also be large
to reflect this. 
In this
toy model $\alpha$ need
only be large enough to ensure that the conditions (\ref{c12Defs}) are satisfied,
so as to 
cause significant distortion of the (eventual) thermal distribution.

The key 
dimensionless parameter is
actually $c_2$, not $\alpha$, as it is the
degree to which $c_2$ exceeds unity that parameterises the importance of the chemical 
potential $\mu$ and controls the 
resulting distribution for high-momentum modes. 
For the chemical potential to be significant for high-momentum modes,
 $c_2 \gtrsim 1$ is necessary. 
 However, this parameter too is 
 constrained by the requirement to capture the 
 post-preheating physics; extremely large values ($c_2 \gg 1$) 
 would lead to an unnaturally sharp
 cutoff in the distribution function and should not be considered.

An alternative way to parameterise the distortion of the number density for high-momentum fermions
would be to employ a momentum-dependent chemical potential $\mu(p)$. The form of $\mu(p)$ should satisfy the following properties:
\begin{itemize}

\item As thermalisation occurs, $\mu(p)\rightarrow 0$. 

\item $\mu(p)<0$ in order to enhance the exponential part of the Fermi-Dirac distribution.

\item $|\mu(p)|$ is an increasing function of $p$, so that the suppression of the number density is enhanced for large momentum modes. 

\item To avoid deviations from the thermal distribution for low-momentum fermions, $p$ should
not become positive for small $p$. 

\end{itemize}

The effect of such a function would be closer to that of $\mu$-distortion in the CMB. 
An example of a function that satisfies these criteria and gives rise to the power law decay $f(p)\sim p^{-4}$ for high-momentum modes characteristic for strong turbulence (as reported in \cite{Berges:2010zv}) is
\eq
\mu(p) = \Bigg\lbrace \begin{array}{cc}
0, & p < p_0 \\
\alpha [-4\beta^{-1}\log ( e^{\tfrac{1}{4}\beta E(p_0)} p/p_0 )+E(p)], & p > p_0
 \end{array}
\eqe 
where $p_0$ is the point at which the suppression takes effect and thermalisation occurs in 
the limit $\alpha\rightarrow 0$. For the remainder of this work, however, we shall use the temperature distortion prescription (\ref{TempPrescription}), for which the coefficients in the
effective Lagrangian can be calculated analytically.

\subsection{The quadratic term $a$}

We expand the $\Delta^*\Delta$ term in (\ref{SEffE}) in gradients using
\begin{multline}
 \frac{1}{i\omega_n-\xi(\bf k +i\nabla)}
=
 \frac{1}{i\omega_n-\xi({\bf k})}
+
 \frac{1}{[i\omega_n-\xi({\bf k})]^2}\frac{i\bf k\cdot \nabla}{m} + \\
 +
 \left\lbrace
  \frac{1}{[i\omega_n-\xi({\bf k})]^2}\left(  \frac{-\nabla^2}{2m} \right)
  -   \frac{1}{[i\omega_n-\xi({\bf k})]^3} \left( \frac{\bf k\cdot \nabla}{m}  \right)^2
   \right\rbrace +\dots
   \label{GradExp}
\end{multline}

The term proportional to $({\bf k\cdot \nabla}) \Delta^*\Delta$ will not contribute as we integrate over all $\bf k$.
For the quadratic term we have
\eq
a=\frac{1}{4\pi}\int_0^{\sqrt{2m\mu}}\frac{dk k}{\xi}\tanh(\beta\xi/2)
+\frac{1}{4\pi}\int_{\sqrt{2m\mu}}^\infty \frac{dk k}{\xi}\tanh(\beta\xi/2)
-\frac{1}{\lambda_0},
\eqe
where we have split the integral at the point $\xi=k^2/2m-\mu=0$. Following the standard BCS treatment, we swap the coupling constant for the binding energy of a fermion pair $\epsilon_a$. The two quantities are related by
\eq
\frac{1}{\lambda_0} = \tfrac{1}{2}\nu(0)\ln\left(\frac{2\epsilon_\Lambda}{\epsilon_a}\right),
\eqe
where $\epsilon_\Lambda = \Lambda^2/2m$ ($\Lambda$ is a momentum cutoff) and $\nu(0)=m/2\pi$
is the 2d density of states per spin degree of freedom.
Next, we apply the temperature prescription (\ref{TempPrescription})
and use the substitution
\eq\label{xSubstitution}
x = \Bigg\lbrace \begin{array}{cc} \beta\xi/(2\alpha) & \xi<0 \\ \alpha\beta\xi/2 & \xi>0    \end{array}
\eqe
where $x$ is unrelated to the spacetime coordinates. This
 gives
\[
a = \frac{1}{2}\nu(0)\left\lbrace 
\int_{-c_1}^0 \frac{X}{x} dx + \int_0^{\alpha\beta\epsilon_\Lambda/2}\frac{X}{x} dx
+\ln\left( \frac{\epsilon_a}{2\epsilon_\Lambda}\right)
 \right\rbrace,
\]
where we define $X=\tanh(x)$ and $c_1$ and $c_2$ are defined in (\ref{c12Defs}).
Since $c_1\ll 1$ we can neglect the contribution
from the first integral.
Making use of (\ref{tanhInt}) with $\epsilon_\Lambda\rightarrow 
\infty$ the second integral can be rewritten
\[
\int_0^{\alpha\beta\epsilon_\Lambda/2} \frac{X}{x} dx \rightarrow \ln\left(\frac{2\alpha\beta\epsilon_\Lambda e^\gamma}{\pi}\right),
\]
so
\eq
a\simeq\tfrac{1}{2}\nu(0)\ln\left( \frac{T_0}{T} \right),
\eqe
with
\eq \label{TransitionTemperature}
T_0\equiv  \frac{\alpha\epsilon_a e^\gamma}{\pi}.
\eqe
This takes the same form as the standard BCS formula but with a different transition temperature
(the standard case has $T_0 = e^\gamma \sqrt{2\mu\epsilon_a}/\pi$).

\subsection{The quartic term $b$}

Using (\ref{Matsubara2}), the quartic term is found to be
\eq
b = -\frac{1}{4}\int \frac{d^2k}{(2\pi)^2} \frac{1}{\xi}\frac{d}{d\xi}\left(\frac{\tanh(\beta\xi/2)}{\xi}\right).
\eqe
Splitting the integral into two parts and using (\ref{TempPrescription})
and (\ref{xSubstitution}) gives
\[
b = -\frac{\nu(0)\beta^2}{16} \left\lbrace  \frac{1}{\alpha^2}
\int_{-c_1}^0 \frac{dx}{x} \frac{d}{dx}\left( \frac{X}{x} \right)   +
\alpha^2  \int_{0}^{\infty} \frac{dx}{x} \frac{d}{dx}\left( \frac{X}{x} \right) 
\right\rbrace.
\]
Neglecting the first integral as before and using (\ref{dtanhInt})
we find 
\eq
b \simeq \frac{7\nu(0)\alpha^2\beta^2}{16\pi^2}\zeta(3).
\eqe

\subsection{The gradient term}

From (\ref{GradExp}) we see that the gradient term has two contributions. Performing the sum using (\ref{Matsubara3}), the first can be rewritten
\eq
-\frac{1}{8m}\int \frac{d^2k}{(2\pi)^2}\frac{d}{d\xi}\left(\frac{\tanh(\beta\xi/2)}{\xi}\right)
 =
-\frac{\nu(0)\alpha\beta}{16m} \left[  \frac{1}{\alpha^2} \int_{-c_1}^\infty dx \frac{d}{dx}\left(\frac{X}{x}\right)
+\int_{0}^\infty dx \frac{d}{dx}\left(\frac{X}{x}\right) 
\right].
\nn
\eqe
Similarly, 
using (\ref{Matsubara4}), the second integral is
\eq
\begin{split}
& -\frac{1}{32\pi m^2}  \int_{-\beta\mu/2}^\infty  dk k^3 \Bigg[  
\frac{\tanh(\beta\xi/2)}{\xi^3}  + \frac{\beta}{2}\frac{d}{d\xi}\left(\frac{\cosh(\beta\xi/2)}{\xi}\right)
\Bigg] \nn\\
& \quad=  -\frac{\nu(0)\alpha\beta}{16m} \left[  \frac{1}{\alpha^2} \int_{-c_1}^\infty dx  \left(x+c_1\right) \left( \frac{X}{x^3}  +\frac{d}{dx}\left( \frac{Y}{x} \right)  \right)+
\int_{0}^\infty dx  \left(x+c_2\right) \left( \frac{X}{x^3}  +\frac{d}{dx}\left( \frac{Y}{x} \right)  \right) \right],
\end{split}
\eqe
where $Y\equiv\mathrm{sech}^2(x)$. Since $X/x^3+(Y/x)'$ is smooth at $x=0$ we can let $c_1\rightarrow 0$ in the integrand and denominator. Combining the two parts gives the gradient term as
\eq
\begin{split}
c  & \simeq  -\frac{\nu(0)\alpha\beta}{16m} \left\lbrace
\int_0^\infty dx \left( \frac{X}{x^2} - \frac{Y}{x} \right) + \left[ \frac{X}{x} + Y \right]^\infty_0 \right\rbrace, \\
& = \frac{\nu(0)\alpha\beta}{16m}.
\end{split}
\eqe

\subsection{Time-dependence}{\label{NonRelTime}}

To take account of the time-dependence, we must return to the quadratic term (\ref{NonRel_S2}), using $\tilde k_0 = i\omega_n +\p_\tau =  i\omega_n -  i\omega_\ell $.
The dynamical part of the action is then found by expanding 
\[
Q(i\omega_\ell ) = 
S_{\rm eff }^{(2)}(\vec k = 0, k_0 = i\omega_n, \tilde k_0 = i\omega_n -  i\omega_\ell  ) -
S_{\rm eff }^{(2)}(\vec k = 0, k_0 = i\omega_n, \tilde k_0 = i\omega_n  ) ,
\]
after analytic continuation to the real axis with the prescription $i\omega_\ell = q_0 +i\epsilon$
\cite{SadeMelo:1993zz}.
The dynamical part of the effective Lagrangian is then given by \cite{Schakel:1999pa}
\eq
\mathcal{L}_{\rm dyn} = [Q'(i\p_0)-i\pi Q''(i\p_0)]\Delta^* \Delta,
\eqe
with
\eq
Q'(q_0) = -\dashint_{\bf k} \frac{q_0\tanh(\beta\xi/2)}{2\xi(2\xi+q_0)}, \qquad
Q''(q_0) =  -\int_{\bf k} \delta(2\xi + q_0)\tanh(\tfrac{1}{2}\beta\xi),
\label{nonRel_QDefs}
\eqe
where $\dashint$ stands for the principal part (the $Q'$ integral has a single pole at $\xi=q_0/2$).
Expanding the integrand about $q_0=0$ gives
\eq
\begin{split}
Q'(q_0)  & = -\frac{\nu(0)}{2} \Bigg\lbrace
\frac{\alpha\beta q_0}{4} 
\left[
\frac{1}{\alpha^2} \ddashint_{-c_1}^0 \frac{X}{x^2} dx + 
\ddashint_0^\infty \frac{X}{x^2} dx 
\right]  + \mathcal{O}(q_0^2) \Bigg\rbrace 
\\
& \simeq
-\frac{\nu(0)\alpha\beta q_0}{8}
\left[
\eta-\frac{\ln(c_1)}{\alpha^2}
\right]
 \\
& \simeq -\frac{\nu(0)\alpha\beta\eta q_0}{8},
\end{split}
\eqe
where in the second line
\eq
\eta\equiv 1 + 2\int_0^\infty XY\ln(x) dx \approx 0.7905,
\eqe
is a constant and 
 the principal value integrals have been evaluated using (\ref{Hypersingx2}).
 In the third line, we have neglected the subdominant term
 $ \alpha^{-2} \ln(c_1) =  \alpha^{-2} \ln(c_2 \alpha^{-2})  \ll 1$
 as $\alpha$ is large.
 The expansion to second order in $q_0$ is performed in Appendix \ref{TimeDerivativeSecondOrder}.
 
The delta function in
(\ref{nonRel_QDefs})
picks out $\xi = -\tfrac{1}{2}q_0$, which, as $q_0 \ll \mu$, is extremely close to $\xi=0$ where the temperature jumps according to the prescription (\ref{TempPrescription}). To avoid unphysical features arising from the treatment of the jump as a step-function we take $\beta(k)=\beta$ here, which gives
\eq\label{NonRelTimeTanhTerm}
Q''(q_0) = \tfrac{1}{2}\nu(0)\tanh\left(\frac{\beta q_0}{4} \right) \simeq
\tfrac{1}{8}\nu(0)\beta q_0 .
\eqe
After integrating by parts, the dynamical part of the Lagrangian is then
\eq
\mathcal{L}_{\rm dyn} \simeq\frac{\beta\nu(0) }{8} \Delta^*
\left( \alpha\eta i -\pi \right)\p_0 
 \Delta.
\eqe

\subsection{The effective Lagrangian}

Putting these parts together gives the effective Lagrangian
\eq
\begin{split}
\mathcal{L}_{\rm eff}   =   \tfrac{1}{2}\nu(0) \Bigg\lbrace \Delta^*
\Bigg[ \ln\left( \frac{T_0}{T} \right) + &
\frac{\beta}{4}\left( \alpha\eta i -\pi \right)\p_0
+\frac{\beta\alpha}{8m}\nabla^2\Bigg]\Delta
-\frac{7\beta^2\alpha^2}{16\pi^2}\zeta(3)|\Delta|^4  \Bigg\rbrace. 
\end{split}
\eqe
Considering only the potential terms, one can see from the expression for the transition temperature (\ref{TransitionTemperature}), 
that if $\alpha$ is large enough, $T<T_0$. The quadratic term in the
effective potential will then be negative while the quartic term is positive;
there is then a minimum at
\eq
|\Delta_{\rm min}| = \sqrt{\frac{16\pi^2}{7\alpha^2\beta^2\zeta(3)}\ln\left(\frac{\alpha\beta\epsilon_a e^\gamma}{\pi} \right)},
\eqe
and the value of the potential at the minimum is negative
\eq
V(|\Delta_{\rm min}|) = -\frac{2\nu(0)\pi^2\ln^2\left(\frac{\alpha\beta\epsilon_a e^\gamma}{\pi} \right)}
{7\alpha^2\beta^2\zeta(3)}.
\eqe
Analysis of the time-independent part thus suggests 
that the effect of the non-thermal distribution function ---
implemented quantitatively with $\alpha \gg 1$ ---
is to give rise to a shift in the ground state of the system.

However, the form of the time-derivative terms is unusual.
The interpretation of the real part of the $\Delta^*\p_0\Delta$ term is that high energy fermion pairs can break up as
in the standard BCS model with weak coupling; the imaginary part indicates that $\Delta$ has a propagating part. 
Time-derivative terms with complex coefficients can also occur in the weak coupling limit of the
BCS-BEC crossover, as a result of a starting Lagrangian that is not particle-hole symmetric \cite{SadeMelo:1993zz}. An important difference here is that the real part is subdominant.
To better understand the effect of the time-derivative terms, it is necessary to repeat the analysis with a relativistic Lagrangian:
this will be the subject of the following sections.


\section{Relativistic fermions}\label{RelSec}

In this section, we apply the methods of the previous section to
a 2+1 dimensional model with relativistic fermions described by the Dirac Lagrangian.
Following some preliminary comments on the treatment of fermions in 2+1 dimensions, we explicitly derive the gravitational four-fermion interaction
and use the result to calculate the effective Lagrangian for the pair field.

\subsection{Fermions in 2+1 dimensions}

To treat the case where the fermions are relativistic,
our starting point must be the Dirac Lagrangian.
However, in 2+1 dimensions this comes hand-in-hand with
several nonintuitive features arising from the fact that in odd spacetime
dimensions there are two inequivalent representations of the gamma matrices, 
which can be written in terms of the Pauli matrices as:
 $\gamma^0 = \sigma_3$, $\gamma^1 = i\sigma_1$, $\gamma^2 = \pm i\sigma_2$.
 (Note that the Dirac spinors have only two components.) 
Following \cite{deJesusAnguianoGalicia:2005ta}, we refer to the $+i\sigma_2$ case as the A representation and $-i\sigma_2$ 
case as the B representation. 

The solutions\footnote{
The particle and antiparticle solutions in the A representation, which
satisfy $(i\slashed{\p}_A-m)\psi_A=0$ with $\gamma^2 = +i\sigma_2$,
can be projected out with the operator $\Lambda_\pm = (\pm \slashed{p}_A+m)/2m$.
Here, the corresponding solutions in the B representation, satisfying 
$(i\slashed{\p}_B-m)\phi_B=0$ with $\gamma^2 = -i\sigma_2$, are written 
as $\phi_B = -\sigma_2 \psi_A$, as $\Lambda_+\psi_A$ and $\Lambda_+\psi_B$ correspond 
to particles with opposite spins.
}
 of the Dirac equation in the A and B representations are not independent,
but are related by a parity transform $\mathcal{P}$:  $(\psi_A)^\mathcal{P} = -i\gamma^1\psi_B e^{i\phi_P}$ and $(\psi_B)^\mathcal{P} = -i\gamma^1\psi_A e^{-i\phi_P}$, where $\phi_P$ is a phase,
which converts a spin-up particle (antiparticle) to a spin-down particle (antiparticle) and vice versa
\cite{Shimizu:1984ik}.

Thus in order to describe fermions with both spins, 
we need the (parity invariant) Lagrangian
\eq
\mathcal{L} = \bar\psi_A(i\slashed{\p}_A-m)\psi_A + \bar\psi_B(i\slashed{\p}_A+m)\psi_B,
\label{LABLagrangian}
\eqe
where the slashed notation $\slashed{\p}_A$ in this equation indicates contraction with the $2\times2$ gamma matrices
in the A representation.
This can be written in the form $\mathcal{L} = \bar\psi(i\slashed{\p}-m)\psi$ using the four-component representation
\eq
\gamma^0 = \begin{pmatrix} 
\sigma_3 & 0 \\ 
\SetToWidest{0} & \SetToWidest{-\sigma_3} 
\end{pmatrix},  \qquad
\gamma^1 = \begin{pmatrix} 
i\sigma_1 & 0 \\ 
\SetToWidest{0} & \SetToWidest{-i\sigma_1}
 \end{pmatrix}, \qquad
\gamma^2 = \begin{pmatrix} 
i\sigma_2 & 0 \\ 
\SetToWidest{0} & \SetToWidest{-i\sigma_2}
 \end{pmatrix},
\eqe
and
$\psi = (\psi_A  \ \psi_B)^T$.
Since there are only three gamma matrices, we have sufficient freedom to define
the following matrices
\eq
\gamma^3 = \begin{pmatrix}
0 & I \\ 
\SetToWidest{I} & \SetToWidest{0}
\end{pmatrix}, \qquad
\gamma^5 = i\begin{pmatrix}
0 & I \\ 
\SetToWidest{-I} & \SetToWidest{0}
\end{pmatrix}, 
\eqe
which anticommute with $\gamma^\mu$.
From these we can define the Hermitian matrix
\eq 
\gamma^{35}\equiv i\gamma^3\gamma^5 = 
\left( \begin{matrix} 1 & 0 \\ \ 0 \ & -1 \end{matrix} \right),
\eqe
which commutes with $\gamma^\mu$ and anticommutes with $\gamma^3$, $\gamma^5$.

In order to derive the equivalent Lagrangian to that used in the
non-relativistic case, we shall rewrite the Lagrangian in the Nambu-Gorkov basis 
(cf. \cite{Deng:2006ed}) with a non-zero chemical potential $\mu$ i.e.
\eq\mathcal{L}=
\tfrac{1}{2}\bar\Psi^{NG} 
\begin{pmatrix} 
i\slashed{\p} - m +\gamma^0\mu & 0 \\
0 & i\slashed{\p} - m -\gamma^0\mu 
  \end{pmatrix}
   \Psi^{NG}.
   \label{NGLagrangian1}
\eqe
The Nambu-Gorkov spinor is
\eq
\Psi^{NG} = \begin{pmatrix} \psi \\ \psi^c \end{pmatrix}, \qquad
\bar\Psi^{NG} = ( \bar\psi \quad \bar\psi^c ),
\eqe
where the superscript $c$ indicates the charge conjugate,
given (up to an overall phase) by
\eq
\psi^c = C\bar\psi^T, \qquad \bar\psi^c = -\psi^T C^\dagger,
\qquad C = \gamma^2 e^{i\phi_C \gamma^{35}},
\label{psi_conj_trans}
\eqe
where $\phi_C$ is a  phase.\footnote{
The unusual form of $C$ arises because the solutions of the Dirac equation in the A and B representations describe particles and antiparticles of the same spin, so the charge conjugation operator does not mix
the fields corresponding to the inequivalent representations.
Hence there are two independent phases $\phi_C^A$ and $\phi_C^B$
corresponding to the operations
$(\psi_A)^\mathcal{C} = \gamma^2(\bar\psi_A)^T e^{i\phi_C^A}$ and 
$(\psi_B)^\mathcal{C} = \gamma^2(\bar\psi_B)^T e^{i\phi_C^B}$, 
and only one linear combination can appear as an overall phase in (\ref{psi_conj_trans}).
This has been shown to be an important consideration when considering the bound states in
 quantum electrodynamics in three spacetime dimensions \cite{Burden:1991mb}. 
}

\subsection{Derivation of the interaction term}

To derive the interaction term, 
we will consider minimally coupled fermions in curved space 
in the Vielbein-Einstein-Palatini formulism, in which the spin connection $\bar\omega_\mu{}^I{}_J$
and the vielbein $e^\mu_I$ are taken as independent variables.
The Lagrangian is
\eq \label{InitalLagrangian}
\mathcal{L} = e \left\lbrace  \tfrac{\Mp}{2} e_I^{\mu} e^{\nu J } R_{\mu\nu}{}^I{}_J  
+ \tfrac{i}{2}\left[\bar\psi e^\mu_I \nabla_\mu \psi - (\nabla_\mu \bar\psi) e^\mu_I \gamma^I \psi\right]
-m\bar\psi \psi
\right\rbrace,
\eqe
where $e=|\det(e^{I}_\mu)|$ and $\Mp= (8\pi G_3)^{-1}$ is the reduced
Planck mass in three spacetime dimensions. 
$R_{\mu\nu}{}^I{}_J$ and the covariant derivatives of the fermions are given in terms of the spin connection by
\eq
R_{\mu\nu}{}^I{}_J  = \partial_\nu \bar\omega_\mu{}^I{}_J - \partial_\mu \bar\omega_\nu{}^I{}_J 
+ \bar\omega_\nu{}^I{}_K \bar\omega_\mu{}^K{}_J
-  \bar\omega_\mu{}^I{}_K \bar\omega_\nu{}^K{}_J.
\eqe
and
\eq
\nabla_\mu\psi = \p_\mu \psi + \tfrac{1}{4}\bar\omega_{\mu I J} \gamma^{[I}\gamma^{J]} \psi,
\qquad
\nabla_\mu\bar\psi = \p_\mu \psi - \tfrac{1}{4} \bar\psi  \bar\omega_{\mu I J} \gamma^{[I}\gamma^{J]} .
\eqe

The spin connection can be decomposed into the symmetric Levi-Civita spin-connection $\omega_\mu{}^I{}_J$ (which is torsionfree and metric compatible i.e. $\nabla_\mu \eta_{IJ} = 0$) and an antisymmetric part $D_\mu{}^I{}_J$ as
\[
\bar\omega_\mu{}^I{}_J = \omega_\mu{}^I{}_J + D_\mu{}^I{}_J,
\]
so that the Lagrangian can be be rewritten as $\mathcal{L} = \mathcal{L}_{LC}+\mathcal{L}_{\rm torsion}$,
where $\mathcal{L}_{LC}$ takes the same form as (\ref{InitalLagrangian}), but with covariant derivatives and $R_{\mu\nu}{}^I{}_J$ given purely in terms of the 
Levi-Civita spin connection $\omega_\mu{}^I{}_J$. 
The second term $\mathcal{L}_{\rm torsion}$ contains the torsion-dependent parts of the original gravitational
and fermion Lagrangians and is given explicitly by
\eq
\mathcal{L}_{\rm torsion}
= e\left\lbrace  
 \frac{\Mp}{2} e_I^{\mu} e^{\nu J}\left(
  D_\nu{}^I{}_K D_\mu{}^K{}_J   -  D_\mu{}^I{}_K D_\nu{}^K{}_J \right)
 + D_{\mu I J} A^{\mu I J}
 \right\rbrace,
\eqe
(plus a total divergence) with
\eq
A^{\mu I J} \equiv \tfrac{i}{8}e^{\mu}_K \bar\psi \left\lbrace \gamma^{[I}\gamma^{J]}, \gamma^K  \right\rbrace\psi
= \tfrac{1}{4}e^{\mu}_K \epsilon^{IJK} (\bar\psi \gamma^{35} \psi).
\label{A_def}
\eqe
In the second equality of (\ref{A_def}) we have used the identity $\left\lbrace \gamma^{[I}\gamma^{J]}, \gamma^K  \right\rbrace = 2  \epsilon^{IJK} \gamma^3\gamma^5 = -2i  \epsilon^{IJK}\gamma^{35}$.
We can now find an explicit expression for the antisymmetric part of the spin-connection
by considering the equation of motion obtained from $\mathcal{L}_{\rm torsion}$ \cite{Dadhich:2010xa}.
Defining the tensors
\eq
E_{\mu}{}^\alpha{}_{\beta} \equiv  e^\alpha_I e^K_\beta D_\mu{}^I{}_K,
\qquad \text{and} \qquad
A^{\mu\rho\sigma} \equiv e^\rho_K e^\sigma_L A^{\mu K L},
\eqe
the Lagrangian can be written
\eq
\mathcal{L}_{\rm torsion} =  e  E^{\alpha\beta\gamma}  \left[
 \tfrac{1}{2}\Mp  \left(E_{\beta\gamma\alpha} - g_{\alpha\gamma} E^\mu{}_{\mu\beta}    \right)
  + A_{\alpha\beta\gamma}\right].
  \label{LagrangianD_inE1}
\eqe
Varying with respect to $E^{\alpha\beta\gamma}$ yields
\eq
E_{\beta\gamma\alpha}+E_{\gamma\alpha\beta} - g_{\alpha\gamma} E^{\mu}{}_{\mu\beta}
-g_{\alpha\beta}E^\mu{}_{\gamma\mu} = \tilde A_{\alpha\beta\gamma} ,
\label{E_EOM}
\eqe
where $ \tilde A_{\alpha\beta\gamma} =  -\tfrac{2}{\Mp}A_{\alpha\beta\gamma}$. 
$E^{\alpha\beta\gamma}$ has three independent traces: $c_1 = E_{\gamma\rho}{}^\rho$,
$c_2 = E_{\rho\gamma}{}^\rho$ and $c_3 = E_\rho{}^\rho{}_\gamma$.
Substituting these into (\ref{E_EOM}) and contracting indices with $g^{\alpha\gamma}$ and
$g^{\gamma\beta}$ yields the relation
\[
c_2 = c_3 + 2 \tilde A^\alpha{}_{\beta\alpha},
\]
so (\ref{E_EOM}) can be rewritten 
in terms of $\tilde A_{\alpha\beta\gamma}$ and an arbitrary vector $U_\gamma = c_3$ as
\[
E_{\beta\gamma\alpha}+E_{\gamma\alpha\beta}
=  \tilde A_{\alpha\beta\gamma} + 2g_{\alpha\beta} \tilde A^\rho{}_{\gamma\rho}
+ g_{\alpha\gamma}U_\beta + g_{\alpha\beta}U_\gamma.
\]
Cyclically permuting indices gives two further equations, which combine to give
\eq 
\begin{split}
E_{\alpha\beta\gamma} & = \tfrac{1}{2}( \tilde A_{\gamma\alpha\beta}
+ \tilde A_{\beta\gamma\alpha}  - \tilde A_{\alpha\beta\gamma} ) + 
(g_{\beta\gamma} \tilde A^\rho{}_{\alpha\rho}+g_{\alpha\gamma} \tilde A^\rho{}_{\beta\rho}- g_{\alpha\beta} \tilde A^\rho{}_{\gamma\rho}) 
+ g_{\beta\gamma}U_\alpha, \\
& = \tfrac{1}{2} \tilde A_{\alpha\beta\gamma} + g_{\beta\gamma}U_\alpha,
\end{split}
\eqe
where in the second line we have used the fact that $\tilde A_{\alpha\beta\gamma}$ is 
totally antisymmetric, as can been seen from (\ref{A_def}). Substituting this back into the Lagrangian
(\ref{LagrangianD_inE1})
gives
\eq
\mathcal{L}_{\rm torsion} = -\frac{e}{2\Mp} A^{\alpha\beta\gamma} A_{\alpha\beta\gamma},
\eqe 
where the symmetric part vanishes due to the antisymmetry of $A^{\alpha\beta\gamma}$; as it should, since
it is a gauge artifact \cite{Dadhich:2010xa}.
Substituting the definition of  $A^{\alpha\beta\gamma}$ in (\ref{A_def})
and using the identity $\epsilon_{IJK}\epsilon^{IJK} = 6$
yields the interaction term
\eq
\mathcal{L}_{\rm torsion}  =  -\frac{3e}{16\Mp} (\bar\psi \gamma^{35}\psi)^2 .
\eqe
To make contact with the fermion Lagrangian in the Nambu-Gorkov basis (\ref{NGLagrangian1})
we shall rewrite $\mathcal{L}_{\rm torsion}$ in terms of the 
charge-conjugated fields (\ref{psi_conj_trans})
using the Fierz identities given in Appendix \ref{Fierz_Sec}. Applying (\ref{zExpansion}) gives
\eq \label{LD_FinalForm}
\mathcal{L}_{\rm torsion}  = \frac{\lambda_0}{2}\left[ (\bar\psi \psi^c)(\bar\psi^c \psi)
+  (\bar\psi \gamma^{35} \psi^c)(\bar\psi^c   \gamma^{35} \psi)
+ \tilde\Phi^\dagger \tilde\Phi +  \tilde\Phi^\dagger_\mu \tilde\Phi^\mu 
  \right],
\eqe
where 
\eq \label{lambda0Def}
\lambda_0 = \frac{3}{32\Mp} = \frac{3}{4}\pi G_3,
\eqe
and the doublets $\tilde\Phi$ and $\tilde\Phi^\mu$ are defined in (\ref{tildePhiDef}).
In what follows, for simplicity we shall focus on the effect of the scalar and $\gamma^{35}$ interactions.\footnote{
Note that the term involving the axi-pseudovector $\tilde\Phi^\dagger_\mu \tilde\Phi^\mu $, carrying as it does a 
sum over Lorentz indices, may be reasonably expected to be a higher energy channel. Considering the axi-pseudoscalar $\tilde\Phi^\dagger \tilde\Phi$, defined in (\ref{tildePhiDef}), in isolation leads to a similar form for the gap equation to that found using the scalar or $\gamma^{35}$ channels, however, using all three together leads to interaction terms amongst the various auxiliary fields.
} 

\subsection{Integrating out the fermions}

Following the procedure used in the non-relativistic case, we remove the four-fermion
interaction by inserting the following terms in the Lagrangian, 
\[
-\frac{1}{2\lambda_0}\left\lbrace [\Delta_1 - \lambda_0\bar\psi^c \psi]^\dagger[\Delta_1 - \lambda_0\bar\psi^c \psi]+
[\Delta_2 - \lambda_0\bar\psi^c \gamma^{35}\psi]^\dagger[\Delta_2 - \lambda_0\bar\psi^c   \gamma^{35} \psi] \right\rbrace,
\]
which correspond to multiplying the path integral
by a Gaussian integral in the auxiliary fields, and thus do not affect the equations of motion.
Variation with respect to $\Delta_1^\dagger$ and $\Delta_2^\dagger$ show that $\Delta_1$ and
$\Delta_2$ describe fermion pairs coupled in the scalar and $\gamma^{35}$ channels
\eq
\Delta_1 =  \lambda_0\bar\psi^c \psi, 
\qquad
\Delta_2 = \lambda_0\bar\psi^c \gamma^{35} \psi.
\eqe
Rather than work directly with $\Delta_1$ and $\Delta_2$, we redefine as follows
\eq \label{DeltaADef}
\Delta_A = \tfrac{1}{2}(\Delta_1+\Delta_2), \qquad 
\Delta_B = \tfrac{1}{2}(\Delta_1-\Delta_2),
\eqe
so the interaction Lagrangian is
\eq
\mathcal{L}_{\text{int}}=
\tfrac{1}{2}\bar\Psi^{NG} 
\begin{pmatrix} 
0 & 0 & \Delta_A & 0 \\
0 & 0 & 0 & \Delta_B \\
\Delta_A^\dagger & 0 & 0 & 0 \\
0 & \Delta_B^\dagger & 0 & 0
  \end{pmatrix}
   \Psi^{NG} - \frac{1}{\lambda_0}(|\Delta_A|^2+|\Delta_B|^2),
   \label{NGLagrangian2}
\eqe
where each entry indicates a $2\times 2$ block.
The advantage of this is that the complete Lagrangian may be split into two parts, composed of 
$\psi_A$ and $\psi_B$ spinors respectively, as
\[
\mathcal{L} = \mathcal{L}_A+\mathcal{L}_B = (\mathcal{L}_A^\psi+\mathcal{L}_A^{\text{tree}})
+(\mathcal{L}_B^\psi+\mathcal{L}_B^{\text{tree}})
\]
with
\eqa
\mathcal{L}_A^\psi &=& \tfrac{1}{2}\bar\Psi^{NG}_A
\begin{pmatrix}
i\slashed{\p} - m +\gamma^0\mu & \Delta_A \\
\Delta_A^\dagger & i\slashed{\p} - m -\gamma^0\mu 
\end{pmatrix}
\Psi^{NG}_A ,
\label{LagA}\\
\nn\\
\mathcal{L}_B^\psi &=& \tfrac{1}{2}\bar\Psi^{NG}_B
\begin{pmatrix}
i\slashed{\p} + m +\gamma^0\mu & -\Delta_B \\
-\Delta_B^\dagger & i\slashed{\p} +m -\gamma^0\mu 
\end{pmatrix}
\Psi^{NG}_B ,
\label{LagB}
\eqae
and
\eq
\mathcal{L}_A^{\text{tree}} = -\frac{1}{\lambda_0}|\Delta_A|^2,
\qquad
\mathcal{L}_B^{\text{tree}} = -\frac{1}{\lambda_0}|\Delta_B|^2.
\eqe

 \section{The massless case}
 \label{MasslessCase}
 
 In this section, we specialise to the extreme case of massless fermions. 
  For simplicity, and because we consider only the the very short time-scales
relevant for the preheating process, we do not consider the effect of a curved background.
Preheating occurs when the homogeneous zero mode of the inflaton field is exhibiting oscillations 
about the minimum of its potential, so that the dominant effect of including spacetime curvature is to introduce an additional friction term that damps the oscillations. Whilst this does have an effect on
 the process of parametric resonance used to produce the fermions, on very small timescales, one can treat the background as constant. 
 
 In this case, the covariant derivatives in the Dirac Lagrangian (taken with the Levi-Civita spin connection) reduce to partial derivatives. 
Here, as in (\ref{LABLagrangian}), the $2\times 2$ gamma matrices are 
 $\gamma^0 = \sigma_3$, $\gamma^1 = i\sigma_1$ and $\gamma^2 = i\sigma_2$.

Considering only the $A$ parts\footnote{
In the massless case, the result for $\Delta_B$ will be the same as that for $\Delta_A$,
since only even terms contribute in the gradient expansion.
}, we can integrate out the fermions to 
get
\eq
Z = \int D\Delta_A^\dagger D\Delta_A \exp\left( iS_{\rm eff} - \frac{i}{\lambda_0}\int_x |\Delta_A|^2  \right),
\eqe
where $S_{\rm eff}$ is the one-loop effective
action
\eq
S_{\rm eff} = -i\text{Tr} \ln[K(p,x)] = -i\text{tr} \int_x \int_k e^{ik\cdot x} \ln [K(p,x)] e^{-ik\cdot x}.
\eqe
Anticipating the derivative expansion \cite{Schakel:1999pa}, we write $K(p,x)$ as
\eq
K(p,x) = G_0^{-1}\left[ I - G_0
\left( \begin{smallmatrix}
0 & -\Delta_A \\
-\Delta_A^\dagger & 0
\end{smallmatrix} \right)\right],
\eqe
with
\eq
G_0 = \begin{pmatrix}
G_+ & 0 \\ 0 & G_-
\end{pmatrix},
\qquad
G_\pm = \frac{1}{(p_0 \pm \mu)^2 - p^2}
\begin{pmatrix}
p_0 \pm \mu & -ip_1 - p_2 \\
-ip_1 +p_2 & -p_0 \mp \mu
\end{pmatrix}.
\eqe
Here $p = \sqrt{p_1^2+p_2^2}$.
Dropping the $\log G_0^{-1}$ term, which is independent of $\Delta_A$ and does not
contribute to the effective action,
we can expand as
\eq
S_{\rm eff} = \sum_{\ell = 1}^{\infty} S_{\rm eff}^{(\ell)} 
 = i\text{Tr} \sum_{\ell = 1}^\infty \frac{(-1)^\ell}{\ell}  \begin{pmatrix}
0 & G_+\Delta_A \\
G_- \Delta_A^\dagger & 0 
\end{pmatrix}^\ell .
\eqe
As in the non-relativistic case only the even powers give a non-zero 
contribution.
 For $\ell=2$ we get
\eq
S_{\rm eff}^{(2)}  = i\text{Tr}\left\lbrace 
\frac{(k_0-\mu)(\tilde k_0+\mu)-k_1 \tilde k_1 - k_2 \tilde k_2}{[(k_0-\mu)^2-k^2][(\tilde k_0+\mu)^2-\tilde k^2]}\Delta_A^\dagger \Delta_A
+
\frac{(k_0+\mu)(\tilde k_0-\mu)-k_1 \tilde k_1 - k_2 \tilde k_2}{[(k_0+\mu)^2-k^2][(\tilde k_0-\mu)^2-\tilde k^2]}\Delta_A \Delta_A^\dagger
 \right\rbrace ,
 \label{Seff2Term}
\eqe
where $\tilde k_\mu = (k_0-i\p_0,k_i+i\nabla_i)$.
 An important subtlety here is that the derivatives here act only on the next object on their right, meaning 
 that after expanding in $\tilde k_\mu$, the first derivatives in the first term in (\ref{Seff2Term})
 pick up a minus sign from integration by parts when the term is expressed in the form $\propto \Delta_A^\dagger \partial_\mu\Delta_A$. 
We include the quartic terms without derivatives, so $p_\mu$ can be treated as a c-number,
giving
\eq
S_{\rm eff}^{(4)}  = \frac{i}{2}\text{Tr}\left\lbrace  
\frac{1}{[k_0^2-\xi(k)]^2} + \frac{1}{[k_0^2-\bar\xi(k)]^2}  
 \right\rbrace |\Delta_A|^4.
\eqe
Here we have introduced the useful notation
\eq \label{xiDef}
\xi(k) = k-\mu, \qquad\qquad \bar\xi(k) = k+\mu.
\eqe
The time-independent part of the Euclidean action is then
\begin{multline}
S_{\rm eff}^E = \int_0^\beta d\tau
\int_{\bf x}\int_{\bf k}\beta^{-1} \sum_n \bigg[ \bigg( 
  M_{Q_1}\nabla^2 + M_{Q_2} (\vec k \cdot \vec\nabla)^2   -\frac{1}{\omega_n^2+\xi^2} -\frac{1}{\omega_n^2+\bar\xi^2} \bigg) |\Delta_A|^2 + \\
+\frac{1}{2}\left(
\frac{1}{[\omega_n^2+\xi]^2} + \frac{1}{[\omega_n^2+\bar\xi]^2}  
 \right) |\Delta_A|^4
 \bigg],
 \label{TIEffectiveEuclidAction}
 \end{multline}
where the coefficients $M_L$, $M_{Q_1}$ and $M_{Q_2}$ are defined in
Appendix \ref{KinExpansion}. 
We have not included the linear term (proportional to $i{\bf  k \cdot \nabla} $) since
it vanishes when we integrate over all $\bf k$.

\subsection{Deriving the effective Lagrangian}

In deriving the effective Lagrangian for $\Delta_A$, we proceed along the same lines as in Sec.
\ref{NonRelEffectiveLagrangianDerivation},
applying the temperature prescription (\ref{TempPrescription}) with $\xi = k-\mu$.
Each of the terms in the Lagrangian (\ref{LagrangianAnsatz})
can be split into terms involving only $\xi$ and $\bar\xi$, as can already be seen in (\ref{TIEffectiveEuclidAction}). The latter, involving the combination $k+\mu$, arise from the
antifermions (cf. \cite{OHSAKU_2+1,Nishida:2005ds,He:2007kd}) and, as we shall see, represent only subdominant contributions.

\subsubsection{The quadratic term $a$}\label{Sec:Rel_QuadTerm}
Using the Matsubura sums in Appendix \ref{IntegralsIdentities}, the quadratic term 
is
\eq\begin{split}
a &= \int \frac{d^2k}{(2\pi)^2} \left[ \frac{\tanh(\tfrac{1}{2}\beta\xi)}{\xi} +  \frac{\tanh(\tfrac{1}{2}\beta\bar\xi)}{\bar\xi}  \right] - \frac{1}{\lambda_0} ,\\
& \equiv a_\xi+a_{\bar\xi}  - \frac{1}{\lambda_0} .
\end{split}
\eqe
We can treat the $\xi$ and $\bar\xi$ parts separately. Starting with the first term, we apply the temperature prescription (\ref{TempPrescription}) to get
\eq\begin{split}
a_\xi & = \frac{1}{4\pi}  \left\lbrace \int_0^\mu dk k  \frac{\tanh(\tfrac{\beta}{2\alpha}\xi)}{\xi} +
 \int_\mu^\Lambda dk k  \frac{\tanh(\tfrac{\alpha\beta}{2}\xi)}{\xi} \right\rbrace, \\
 & = \frac{1}{4\pi}  \left\lbrace \frac{2\alpha}{\beta} \int_{-c_1}^0 dx (x+c_1) \frac{X}{x} 
  +  \frac{2}{\alpha\beta}\int_{0}^{\alpha\beta\Lambda/2} dx (x+c_2) \frac{X}{x} 
\right\rbrace,
 \end{split}
\eqe
where $c_1$ and $c_2$ are defined in (\ref{c12Defs}), $\Lambda$ is a UV cutoff, and we have used $x = \beta\xi/(2\alpha)$ in the first integral
and $x = \alpha\beta\xi/2$ in the second. 
In the limit $c_1 \rightarrow 0$ the first integral can be neglected. Thus
\eq\begin{split}
a_\xi & \simeq \frac{1}{2\pi\alpha\beta}\left[ \ln(\cosh(\alpha\beta\Lambda/2)) +c_2 \ln\left( \frac{2\alpha\beta\Lambda e^\gamma}{\pi} \right) \right] , \\
& \simeq \frac{\Lambda}{4\pi} +  \frac{1}{2\pi\alpha\beta}\left[  c_2 \ln\left( \frac{2\alpha\beta\Lambda e^\gamma}{\pi} \right) -\ln 2 \right] .
\end{split}
\eqe
Returning to the $\bar\xi$ integral, we proceed in a similar fashion and split the integral at $k=\mu$. Using the substitutions
$x = \beta\bar\xi/(2\alpha)$ and $x = \alpha\beta\bar\xi/2$ and the definition (\ref{xiDef}) we get
\eq
\begin{split}
a_{\bar\xi} &=  \frac{1}{4\pi}  \left\lbrace \frac{2\alpha}{\beta} \int_{c_1}^{2c_1} dx (x-c_1) \frac{X}{x} 
  +  \frac{2}{\alpha\beta}\int_{2c_2}^{\alpha\beta\Lambda/2} dx (x-c_2) \frac{X}{x} 
\right\rbrace , \\
& \simeq \frac{\Lambda}{4\pi} +  \frac{1}{2\pi\alpha\beta}\left[  -c_2 \ln\left( \frac{2\alpha\beta\Lambda e^\gamma}{\pi} \right) -\ln 2 - \int_0^{2c_2} dx (x-c_2) \frac{X}{x}  \right], \\
& \simeq \frac{\Lambda}{4\pi} +  \frac{1}{2\pi\alpha\beta}\left[  -c_2 \ln\left( \frac{2\alpha\beta\Lambda e^\gamma}{\pi} \right) + c_2\ln\left( \frac{4e^{\gamma-2} c_2}{\pi} \right)  \right],
\end{split}
\eqe
where, as before, we have neglected the first integral in the limit $c_1\rightarrow 0$.
In the last line we used the fact that $c_2\gtrsim 1$ to simplify the integral using $\int_0^{2c_2} Xdx \simeq 2c_2 -\ln 2$ and (\ref{tanhInt}).
The quadratic
function can then be written
\eq\label{aTermRel}
a\simeq \frac{\Lambda}{2\pi}  +  \frac{1}{2\pi\alpha\beta}\left[ c_2\ln\left( \frac{4e^{\gamma-2} c_2}{\pi} \right) -\ln 2  \right] -\frac{1}{\lambda_0}.
\eqe
In the nonrelativistic calculation, the explicit dependence on the cutoff was removed
by expressing the coupling constant $\lambda_0$ in terms of the
binding energy in vacuum, obtained by solving the Schr\"odinger equation
for a fermionic pair in vacuum.
Since in this case we deal with massless fermions, this cannot be done.
Nevertheless, it is instructive to consider the equivalent problem for relativistic bound states
using the instantaneous Bethe-Salpeter equations.
For spinless fermion-fermion bound states with a scalar interaction, the scalar part of the reduced wavefunction $\Psi_S$ satisfies \cite{Suttorp:1979sd}
\eq
\frac{E}{k^2}(E^2-\epsilon_B^2)\Psi_S(k) = \int_{\vec k'} V_S \Psi_S(\vec k'),
\eqe
where $E = k^2+m^2$, and the bound state has a mass $2\epsilon_B<2m$.  As in the nonrelativistic case, we consider for simplicity an attractive scalar contact interaction $\lambda_0\delta(\vec x)$
(cf. \cite{Schakel:1999pa})
 which gives
\eq
\frac{1}{\lambda_0} = \frac{1}{2\pi}\int_0^{\Lambda} dk \frac{k^3}{\sqrt{m^2+k^2}(m^2+k^2-\epsilon_B^2)}
\simeq
\frac{\Lambda}{2\pi} - \frac{m}{2\pi}\left( 1+ \frac{1-B^2}{B}{\rm arctanh}(B) \right) +\mathcal{O}(\Lambda^{-1}),
\eqe
where $B=\epsilon_B/m$. In the limit of small $m$ with $B$ finite we see the same dependence on the
cutoff as exhibited in (\ref{aTermRel}), with corrections of $\mathcal{O}(m)$.
This suggests that (for $c_2 \gtrsim 1$) the leading term in $a$ is the logarithm and the coefficient is approximately given by
\eq
a \approx \frac{c_2}{2\pi\alpha\beta}\ln\left( \frac{4e^{\gamma-2} c_2}{\pi} \right).
\label{rel_quadTerm}
\eqe
We postpone a discussion on the validity of this approach until Sec. \ref{Discussion}.

\subsubsection{The quartic term $b$}

Using (\ref{Matsubara2}), the quartic term is
\eq
b = -\frac{1}{4} \int \frac{d^2 k}{(2\pi)^2}\left[ 
\frac{1}{\xi}\frac{d}{d\xi} \left( \frac{\tanh(\tfrac{1}{2}\beta \xi)}{\xi} \right)
+ \frac{1}{\bar\xi}\frac{d}{d\bar\xi} \left( \frac{\tanh(\tfrac{1}{2}\beta \bar\xi)}{\xi} \right)
 \right].
\eqe
Repeating the steps taken in the evaluation of $a$, we obtain 
\eq\begin{split}
b &= -\frac{\alpha\beta}{16\pi} \left\lbrace 
\int_0^\infty dx \left(\frac{x+c_2}{x}\right)\frac{d}{dx}\left( \frac{X}{x} \right)  
+
\int_{2c_2}^\infty dx \left(\frac{x-c_2}{x}\right)\frac{d}{dx}\left( \frac{X}{x} \right)  
\right\rbrace , \\
& =  \frac{\alpha\beta}{16\pi} \left\lbrace 
1+\frac{\tanh(2c_2)}{2c_2} - c_2\int_0^{2c_2} \frac{dx}{x} \frac{d}{dx}\left( \frac{X}{x} \right) 
\right\rbrace , \\
& =  \frac{\alpha\beta}{16\pi} \left\lbrace 
1+\frac{\tanh(2c_2)}{8c_2} +\frac{\sech^2(2c_2)}{4} + c_2 \int_0^{2c_2} dx \frac{XY}{x} 
\right\rbrace , \\
& \approx  \frac{\alpha\beta}{16\pi} \left\lbrace 
1+
\frac{7\zeta(3)}{\pi^2} c_2
\right\rbrace ,
\end{split}
\label{rel_QuartTerm}
\eqe
where in the third line we have used (\ref{App_dDx_x}). 
In the fourth line, we have neglected the subdominant $\tanh$ term (which is $< 0.12$ for $c_2>1$) and used (\ref{XYx}) and (\ref{XYxApprox}).

\subsubsection{The gradient term}

From (\ref{TIEffectiveEuclidAction}) it can be seen that there are two contributions
to the coefficient of the gradient term in the Lagrangian. (The linear term $M_L $ need not
be considered as it vanishes when we perform the angular integral.)
Taking account of the factor of $\tfrac{1}{2}$ arising from the $\cos^2\theta$ in the scalar 
product, the coefficient is
\[
M = -\sum_n \left[ M_{Q_1} + \tfrac{1}{2}k^2 M_{Q_2} \right] = M_\xi + M_{\bar\xi} \ ,
\]
where the coefficients $M_{Q_1}$ and $M_{Q_2}$ are given by (\ref{RelKineticTerms}) and
\eq\begin{split}
M_\xi  &= \frac{1}{64\pi}\int_0^\infty
dk \left\lbrace
\frac{\beta^2 k}{\xi} \sech^2(\tfrac{1}{2}\beta\xi)\tanh(\tfrac{1}{2}\beta\xi)
+\beta\mu\frac{\sech^2(\tfrac{1}{2}\beta\xi)}{\xi^2}
+\frac{2(\xi-\mu)}{\mu\xi^3}\tanh(\tfrac{1}{2}\beta\xi)
\right\rbrace , \\
& = 
\frac{\beta}{64\pi\alpha} \int_{-c_1}^0 dx \tilde M_\xi 
+ \frac{\alpha\beta}{64\pi}  \int_0^\infty dx \tilde M_\xi.
\end{split}
\eqe
In the first line we have evaluated the Matsubura sums using (\ref{KineticTermMatSums})
and in the second applied the prescription (\ref{TempPrescription}). 
Similarly, the second term is
\eq
 M_{\bar\xi} =\frac{\beta}{64\pi\alpha} \int_{c_1}^{2c_1} dx \tilde M_{\bar\xi} 
+ \frac{\alpha\beta}{64\pi}  \int_{2c_2}^\infty dx \tilde M_{\bar\xi} \ , 
\eqe
where the integrands are given by
\begin{subequations}
\begin{align}
\tilde M_\xi & = \frac{2(x+c_2)}{x}XY +c_2\frac{Y}{x^2} +
\left( \frac{x^2-c_2^2}{c_2} \right) \frac{X}{x^3}, \\
\tilde M_{\bar\xi} & = \frac{2(x-c_2)}{x}XY -c_2\frac{Y}{x^2} -
\left( \frac{x^2-c_2^2}{c_2} \right) \frac{X}{x^3},
\end{align}
\end{subequations}
and we have evaluated the Matsubura sums using (\ref{KineticTermMatSums}).
Neglecting the $c_1$ integrals\footnote{
Unlike the previous two cases, to see this one has to consider both 
both the $\xi$ and $\bar\xi$ parts. Replacing the integrands by their Maclaurin series
$\tilde M_{\xi,\bar\xi}  = \pm\frac{4c_1}{3}\pm\frac{1}{c_1}+2x\mp \left(\frac{32}{15}c_1+\frac{1}{3c_1}\right)x^2 +\mathcal{O}(x^3)$
 for small $c_1$ one has
\[
\int_{-c_1}^0 dx \tilde M_\xi +  \int_{c_1}^{2c_1} dx \tilde M_{\bar\xi} = \tfrac{8}{3}c_1^2+\mathcal{O}(c_1^4),
\]
which can be neglected as $c_1\rightarrow 0$.
}
we find
\eq
\begin{split}
M &=  \frac{\alpha\beta}{64\pi} \left[ 2 \hspace{0.1em} \sech^2(2c_2)+   \int_0^{2c_2} dx \tilde M_\xi \right], \\
&= \frac{\alpha\beta}{64\pi} \left[
1+ \frac{1}{8c_2}\left[ \tanh(2c)+6\sech^2(2c_2) \right]
+ \frac{1}{c_2}\int_0^{2c_2} dx (c_2^2 Y +1)\frac{X}{x}
\right] ,
\end{split}
\eqe
where in the second line we have used (\ref{XxY}).
We can make use of use of (\ref{tanhInt}) and (\ref{XYx}) to obtain the following approximation
\eq
M \simeq \frac{\alpha\beta}{64\pi} \left[  
1+ \frac{7\zeta(3)}{\pi^2} c_2 +\frac{1}{c_2} \ln\left( \frac{8c_2 e^\gamma}{\pi}  \right)
\right],
\label{rel_GradTerm}
\eqe
which is valid for $c_2\gtrsim 1$.

\subsubsection{Time-dependence}

Following the procedure discussed in Sec. \ref{NonRelTime},
the time-dependent part of the effective Lagrangian is 
\eq
\mathcal{L}_{\rm dyn} = [Q_{(+)}'(i\p_0) - i\pi Q_{(+)}''(i\p_0)]\Delta^\dagger_A \Delta_A
+[   Q_{(-)}'(i\p_0) - i\pi Q_{(-)}''(i\p_0)] \Delta_A \Delta^\dagger_A,
\label{Rel_LDyn1}
\eqe
with
\begin{subequations}
\begin{align}
Q_{(\pm)}'(q_0) &= -\frac{q_0}{4} \left( \dashint_{\vec k} \frac{\tanh(\tfrac{1}{2}\beta\xi)}{\xi(q_0\pm2\xi)}
+  \dashint_{\vec k} \frac{\tanh(\tfrac{1}{2}\beta\bar\xi)}{\bar\xi(q_0\mp2\bar\xi)}
\right) ,
\label{QDashRel} \\
Q_{(\pm)}''(q_0) &= \mp \tfrac{1}{2}\int_{\vec k} \left[ \tanh(\tfrac{1}{2}\beta\xi)\delta(q_0\pm2\xi)
- \tanh(\tfrac{1}{2}\beta\bar\xi)\delta(q_0\mp2\bar\xi)\right] .
\label{QDDashRel}
\end{align}
\end{subequations}
The barred integral indicates that the principal part should be computed. 
Expanding (\ref{QDashRel}) in $q_0$ to first order and applying the prescription (\ref{TempPrescription}) gives
\eq
\begin{split}
Q_{(-)}'(q_0) = -Q_{(+)}' &= \frac{q_0}{8}\left[
\ddashint_{\vec k} \frac{\tanh(\tfrac{1}{2}\beta\xi)}{\xi^2} -
\ddashint_{\vec k} \frac{\tanh(\tfrac{1}{2}\beta\bar\xi)}{\bar\xi^2} 
  \right], \\
  &=  \frac{q_0}{16\pi}\bigg[
  \ddashint_{-c_1}^0\left( \frac{x+c_1}{x} \right)X dx
  - \int_{c_1}^{2c_1}  \left( \frac{x-c_1}{x} \right)X dx \ + \\
  & \qquad\qquad\qquad\qquad
  + \ddashint_0^\infty \frac{X}{x^2} dx
  + \int_0^{2c_2} \frac{X}{x} dx
  + c_2 \int_{2c_2}^\infty \frac{X}{x^2} dx
  \bigg],
\end{split}
\eqe
where in the second line we have applied the prescription (\ref{TempPrescription}) and converted to $x$ as before,
combining integrals where possible. Since, using (\ref{Hypersingx2b}) we have
\[
c_1\ddashint_{-c_1}^0 \frac{X}{x^2} dx \approx -c_1\ln(c_1) ,
\]
and the other $c_1$ integrals are $\mathcal{O}(c_1)$, we can neglect their contribution. 
Using (\ref{tanhInt}), (\ref{Hypersingx2a}) and (\ref{Xx2AsymptoticSeries}), when $c_2\gtrsim 1$ we get
\eq
Q_{(-)}'(q_0) = -Q_{(+)}'  \simeq \frac{q_0}{16\pi}\left[  
\eta c_2 +\ln\left( \frac{8e^{\gamma+1/2}c_2}{\pi} \right) \right].
\label{Q'+-Terms}
\eqe

Turning to the $Q''$ terms, we see from (\ref{QDDashRel}) that the 
second delta functions pick out $k = -\mu+\tfrac{1}{2}q_0$ and $k = -\mu-\tfrac{1}{2}q_0$.
As $q_0 \ll \mu$, both of these are always negative so these terms vanish.
The first delta function picks out
$k = \mu-\tfrac{1}{2}q_0$ in $Q_{(+)}''(q_0)$ and $k=\mu+\tfrac{1}{2}q_0$ in $Q_{(-)}''(q_0)$.
As in the nonrelativistic case, by the nature of the expansion $q_0 \ll \mu $ and these values are extremely tiny shifts in $k$ on either side of the temperature jump at $k=\mu$. Thus,
to avoid
an unphysical artefact of the step function prescription,
we evaluate $1/T$ at the midpoint $1/T = \beta$, giving
\begin{subequations}\label{Q''+-Terms}
\begin{align}
Q_{(+)}''(q_0) &= \frac{1}{4\pi}(\mu-\tfrac{1}{2}q_0)\tanh\left(\frac{\beta q_0}{4}\right)
\simeq  \frac{\beta\mu}{16\pi}q_0 +\mathcal{O}(q_0^2), \\
Q_{(-)}''(q_0) &= \frac{1}{4\pi}(\mu-\tfrac{1}{2}q_0)\tanh\left(\frac{\beta q_0}{4}\right)
\simeq  \frac{\beta\mu}{16\pi}q_0 +\mathcal{O}(q_0^2).
\end{align}
\end{subequations}
Substituting (\ref{Q'+-Terms}) and (\ref{Q''+-Terms}) into (\ref{Rel_LDyn1}), and integrating the $\Delta_A^\dagger \Delta_A$ term by parts, we find that the $Q_{(\pm)}''$ parts cancel,
giving
\eq
\mathcal{L}_{\rm dyn} \simeq \frac{i}{8\pi}  \Delta_A^\dagger \left[  
\eta c_2 +\ln\left( \frac{8e^{\gamma+1/2}c_2}{\pi} \right) \right] \p_0\Delta_A.
\label{Rel_LDyn2}
\eqe

\subsection{The effective Lagrangian arising from massless fermions }

Combining equations (\ref{rel_quadTerm}), (\ref{rel_QuartTerm}), (\ref{rel_GradTerm}) and (\ref{Rel_LDyn2}) we arrive at the effective Lagrangian
for $\Delta_A$
\eq
\begin{split}
\mathcal{L}_{\rm eff}  \simeq
\Delta_A^\dagger \Bigg\lbrace &
\frac{1}{8\pi}    \left[  
\eta c_2 +\ln\left( \frac{8e^{\gamma+1/2}c_2}{\pi} \right) \right] i\p_0
+ \frac{c_2}{16\pi} \left[  
1+ \frac{7\zeta(3)}{\pi^2} c_2 +\frac{1}{c_2} \ln\left( \frac{8c_2 e^\gamma}{\pi}  \right)
\right]\frac{\nabla^2}{2\mu} + \\
&+\frac{\mu}{4\pi}\ln\left( \frac{4e^{\gamma-2} c_2}{\pi} \right)
\Bigg\rbrace
\Delta_A
-  \frac{c_2}{16\pi} \left[
1+
\frac{7\zeta(3)}{\pi^2} c_2
\right] \frac{1}{\mu} |\Delta_A|^4.
\end{split}
\label{EffectiveLagrUnsimplified}
\eqe
We recall at this point that, from Eqns. (\ref{lambda0Def}) and (\ref{DeltaADef}),
$\Delta_A$ is related to the Planck mass by
\eq
\Delta_A =\frac{3}{64\Mp}(\bar\psi^c \psi+\psi^c \gamma^{35} \psi).
\eqe
The value of the pair field is then necessarily small, which is consistent with our approach 
(as in neglecting the time derivatives in the quartic term, we have implicitly assumed $q_0\gg |\Delta_A|$.) 
The Lagrangian (\ref{EffectiveLagrUnsimplified})
may be compared with the time-dependent Ginzburg-Landau theory \cite{Abrahams:1966zz,Schakel:1999pa}, also given in terms of an auxiliary field consisting of
a fermion bilinear multiplied by a weak, dimensionful coupling parameter. In that case, the
effective Lagrangian depends on the fermion mass, the chemical potential and the coupling strength, the latter entering
the effective Lagrangian only via the 
 transition temperature in the quadratic
term (which involves the expression for the binding energy). Here, the entire Lagrangian depends only on one dimensionful parameter $\mu$,  
which is related to the
inflaton mass (and thus should be much smaller than the Planck scale). 

\begin{figure}[t]
\centering 
\includegraphics[height=5.7cm]{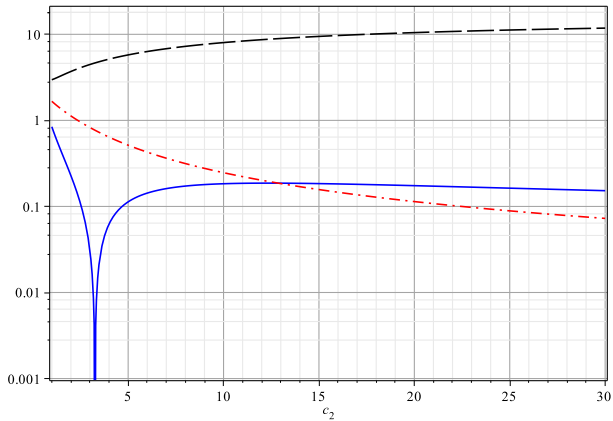}
 \caption{
The combinations $\hat\mu/\mu$ (solid, blue), $\hat m/\mu$ (dot-dash, red) and $\hat\lambda \mu$ (dashed, black) as a function of $c_2$.  $\hat\mu/\mu$ is negative for $c_2 < \pi e^{2-\gamma}/\pi\approx 3.26$.
} 
\label{Coeff_plot}
\end{figure}

For fixed temperature, $c_2$ is a dimensionless constant so we can perform the
field redefinition
\eq
\hat\Delta_A = \left( \frac{1}{8\pi}    \left[  
\eta c_2 +\ln\left( \frac{8e^{\gamma+1/2}c_2}{\pi} \right) \right] \right)^{1/2} \Delta_A,
\eqe
involving the numerical factors $\eta\approx 0.79$ and $c_2\gtrsim 1$, so that 
\eq
\mathcal{L}_{\rm eff}  \simeq \hat \Delta_A^\dagger \Bigg\lbrace
i\p_0 + \hat\mu+ \frac{\nabla^2}{2\hat m}
\Bigg\rbrace
\hat\Delta_A
-\hat\lambda |\hat\Delta_A|^4,
\label{RelCase_Final_L}
\eqe
where
\begin{align}
\hat m  &\equiv 2\mu \eta \left[1+\frac{\ln\left( 8c_2 e^{\gamma+1/2}/\pi \right)}{\eta c_2}\right]\left[1+ \frac{7\zeta(3)}{\pi^2} c_2 +\frac{1}{c_2} \ln\left( \frac{8c_2 e^\gamma}{\pi}  \right)\right]^{-1} , \\
\hat\mu &\equiv  2\mu \ln\left( \frac{4e^{\gamma-2} c_2}{\pi} \right)\left[  
\eta c_2 +\ln\left( \frac{8e^{\gamma+1/2}c_2}{\pi} \right) \right]^{-1}, \\
\hat\lambda &\equiv \frac{4\pi c_2}{\mu} \left[
1+
\frac{7\zeta(3)}{\pi^2} c_2
\right] \left[  
\eta c_2 +\ln\left( \frac{8e^{\gamma+1/2}c_2}{\pi} \right) \right]^{-2}.
\end{align} 
In this form we see that the effective theory for the auxiliary field $\Delta_A$
is that of a weakly interacting Bose gas, described by the Gross-Pitaevski theory 
(see \cite{Andersen:2003qj} for a review).
The mass of the bosons is of the order of the mass scale $\mu$, which is comparable to the inflaton mass. The dependence of these coefficients on the value of $c_2$ is shown in Fig. \ref{Coeff_plot}. 
It is important to note that for $c\gtrsim 3.26$, $\hat \mu$ is positive, allowing for the formation of a
condensate. 

In the strong coupling limit of the BCS-BEC crossover in 2+1 dimensions, 
one finds a similar Gross-Pitaevski equation \cite{DrechslerZwerger}.
In that case the effective chemical potential $\hat \mu$ is a function of the binding energy
but the repulsive interaction depends only on the density of states, 
reflecting the statistical interaction due to the Pauli principle. 
What is striking about our case is that this form arises even though the 
gravitational coupling is extremely weak. 
Since we work with massless fermions, 
(\ref{RelCase_Final_L}) is independent of any binding energy between the 
fermions: the formation of the condensate is determined only by the
magnitude of the departure from thermal equilibrium, as parameterised by $c_2$.  

While the final result may take a simple form, it is not immediately obvious that the effective 
Lagrangian for $\Delta_A$ would describe a propagating field. The presence of an extremely weak coupling and an effective low-temperature limit might lead one to expect BCS condensate behaviour 
(as discussed in other studies involving condensates with the gravitational four-fermion interaction term)
a characteristic feature of which is the breaking up of Cooper pairs at high energy. In this case, the opposite happens: the fermions pairs can be treated as propagating bosons even at high energies. It is the combination of the temperature ansatz (that renders low-momentum modes irrelevant) and the
absence of a tight cutoff in momentum space (a consequence of the gravitational nature of the
interaction term) that  give rise to this behaviour in spite of the BCS-like starting point. 

\subsection{Heuristic explanation}

In both the nonrelativistic and massless cases the quadratic and quartic potential terms
in the effective Lagrangian for the pair field have appropriate signs for
the formation of a condensate. This is possible partly because the 
 the four-fermion interaction
is attractive, which gives the correct sign for the bare quadratic term,
and partly because the nonthermal distribution generates conditions equivalent to the presence of a Fermi surface for high-momentum modes i.e. a nonnegligible 
chemical potential and low effective temperature.
The latter is important both because it prevents the range of integration (which depends on the combination $\mu\beta$) used in deriving the coefficients
from vanishing in the high temperature limit and because the transition temperature
that enters the logarithm in the quadratic term is enhanced by a factor $\alpha$. 
Heuristically, this is a result of the fact that the interaction is not confined to a small region in
momentum space as in the standard BCS case  --- i.e. it is a contact interaction --- so the contribution of high-momentum modes is important. 

It is also important to emphasise here that the torsion mediated interaction does not
introduce a new coupling constant: the dependence of $\lambda_0$ on the Planck mass arises purely as a result of considering Einstein-Cartan gravity. The fermions themselves source the (non-propagating) torsion, as they have intrinsic spin.

The kinetic terms of the two examples differ: the massless fermion condensate has a
purely propagating mode while the nonrelativistic case has a small, real, dissipative part.
This 
would be dominant in the standard BCS case; here the asymmetry between the effective temperature felt by the
modes on either side of the Fermi surface gives rise to an additional imaginary term.
Although one also obtains a Gross-Pitaevski equation in the strong coupling
limit of the BCS-BEC crossover, there these terms are
absent because the chemical potential is negative,
 and the tightly-bound pairs cannot beak up. 
The difference arises in this model due to the inclusion of antifermions in the relativistic case, which cancel the
real first order derivative terms entering the Lagrangian.

\section{Discussion}\label{Discussion}

The starting action was (with the addition of the necessary curvature squared terms)
renormalisable, however, in integrating out the antisymmetric part of the connection
we have obtained an apparently nonrenormalisable four-fermion term. 
The status of this term seems a little different from the standard BCS case,
in which the bare coupling constant $\lambda_0$ is renormalised by the considering the
effective action for the fermions up to one-loop order. As this procedure 
could have been performed after renormalisation of the gravitational constant,
 $\lambda_0$ is already written in terms of renormalised quantities.
As mentioned in the introduction, different approaches to this issue have been taken
in the literature. 
A consistent approach would perhaps be to renormalise the entire Lagrangian
(gravitational and fermion) simultaneously, although since the four-fermion
interaction term is naively nonrenormalisable, it is not
clear whether in this case $\lambda$ would simply be determined by the experimentally
observed value. 
Here, despite the fact that integrating out $D_\mu{}^I{}_J$ can be done
exactly, we have 
treated the resulting Lagrangian as an effective field theory valid up to a scale $\Lambda$,
taken to be of the order of the scale defined by the inverse coupling $\lambda_0^{-1}\sim \Mp$

This is important when treating the quadratic term in
the effective Lagrangian for the pair field, since it is this term that represents a
correction to the bare coupling.
In the nonrelativistic calculation, the logarithmic dependence on the cutoff
does not enter the
effective Lagrangian explicitly, as the $\Lambda$ terms cancel when one
 expresses $\lambda_0^{-1}$ 
in terms of the binding energy $\epsilon_a$, calculated using the Schr\"odinger equation.
In the relativistic calculation, since we have treated the fermions as
massless fields, this is not possible. In Sec. \ref{Sec:Rel_QuadTerm}, it
is shown that there is in this case a linear dependence on $\Lambda/2\pi$,
making it difficult to estimate the magnitude of this term.
Consideration of the instantaneous Bethe-Salpeter equations for a spinless fermion-fermion
bound state with a scalar interaction suggests that the combination
that appears in this term, $\lambda_0^{-1}-\Lambda/2\pi$,
should involve only negligible corrections of the order of the fermion mass. The form of the quadratic term $a|\Delta_A|^2$ is then logarithmic, as in the
nonrelativistic case.
For a rigorous treatment, one should include the mass from the beginning, however,
since this introduces complications, we defer this to future work.

The equivalent calculation in the relativistic massive case
would also be interesting because it would introduce another physical
scale, the fermion mass, into the calculation.
As mentioned in the introduction, many authors have focused on
neutrino condensates with a view to relating the neutrino mass scale
to dark energy. The condensation mechanism presented here is not
particular to neutrinos, so one may think that 
if neutrinos can condense, so can other species.
However, neutrinos are unusual in that they are the only known fundamental 
neutral fermions, so there is less competition between the gravitational interaction
and gauge interactions compared to other species (cf. \cite{Barenboim:2010db}).

 The result (\ref{RelCase_Final_L})
takes the form of a Gross-Pitaevski theory describing a weakly interacting composite Bose gas.
In the BCS-BEC crossover using nonrelativistic fermions, one finds
a similar equation in the strong coupling regime  \cite{Schakel:1999pa}, where the bosons have
twice the fermion mass.
In this case, the mass of the boson is of the order of the only mass scale 
in the theory (with the exception of that defined by the coupling $\lambda_0^{-1}\sim \Mp$)
which is directly related to the inflaton mass. Thus,
the mass of the boson in the effective field theory arises from the kinetic energy of the fermions.
In the standard case, the appearance of a purely propagating
mode in the BEC limit is related 
to the spontaneous breaking of the global U(1) symmetry i.e.
$\Delta$ is a (pseudo) Goldstone mode.
To fully understand and interpret the calculation presented here, it is therefore
important to identify the role played by symmetry breaking.
This is also particularly important in order to be able to comment on the
change of vacuum energy of the system.

 In our ansatz, thermalisation corresponds to the limit
$\alpha\rightarrow 1$, at which point the fermions have a thermal distribution
characterised by a high temperature $T=1/\beta$,
such that $\beta\mu$ is negligible. 
It is important to note that one cannot take this limit directly in (\ref{RelCase_Final_L}),
since the assumptions (\ref{c12Defs})
have been repeatedly used in the derivation. 
Since in this calculation
 $\mu$ is merely a scale used to parameterise the nonthermal distribution
felt by the fermions subsequent to their creation in the preheating process,
after thermalisation this should be replaced by a chemical potential term
relating to any excess of fermions over antifermions, which should be zero.\footnote{
Since the pair field consists of fermions, rather than being of the form $\bar\psi \psi$,
the possibility that a small fraction of bosons could survive thermalisation
is particularly interesting. 
If the density of bosons were small, the bound fermions would not
be able to annihilate with their antiparticles in the same way as their free counterparts,
which could give rise to a lepton number violation that could be
converted to a baryon number violation by sphaleron processes in the early universe. 
However, this possibility seems unlikely.
}

Despite the fact that the inflaton field $\phi$ is also coupled to the fermions (perhaps indirectly) since the latter are produced by its decay, in this simplified description we have not included
its effect directly. Although, like the four-fermion interaction (\ref{LD_FinalForm}), 
the Yukawa interaction $h\phi\bar\psi\psi$
is attractive, the effects of the two terms differ as 
$\phi$ acts like a mass term for the fermions, coupling $\bar\psi$ and $\psi$.
To properly treat the thermalisation limit, one would need to include the effect of energy exchange between the fermions and the $\phi$-bosons; however, this is beyond the scope of this work.

To summarise, by considering a simplified model in 2+1 spacetime
dimensions, we have shown for the first time that the nonthermal distribution 
of fermions arising from preheating after inflation can give rise to a 
fermion condensate. By considering the effective Lagrangian for the spacetime
dependent pair field
 in the extreme cases of nonrelativistic
and massless fermions, 
we have shown that 
it describes a gapless, propagating mode.

\section*{Acknowledgements} 
I would like to thank F. R. Klinkhamer and C. Rahmede for helpful discussions 
and comments and J. Berges for permission to reproduce Fig. \ref{f_plot}.  This work is supported
 by the ``Helmholtz Alliance for Astroparticle Physics HAP",  funded by the Initiative and Networking Fund of the Helmholtz Association.
We would also like to thank the anonymous referee for helpful comments.


\appendix

\section{Integrals and sums}
\label{IntegralsIdentities}

The sums over the Matsubara frequencies are
\eqa
\beta^{-1}\sum_n \frac{1}{\omega_n^2+\xi^2}  &=& \frac{\tanh(\beta\xi/2)}{2\xi},
\label{Matsubara1} \\
\beta^{-1}\sum_n  \frac{1}{(\omega_n^2+\xi^2)^2}  &=& -\frac{1}{4\xi}\frac{d}{d\xi}\left(\frac{\tanh(\beta\xi/2)}{\xi}\right), \label{Matsubara2} \\
\beta^{-1}\sum_n  \frac{1}{i\omega_n+\xi}\frac{1}{(i\omega_n-\xi)^2}  &=& -\frac{1}{4}\frac{d}{d\xi}\left(\frac{\tanh(\beta\xi/2)}{\xi}\right), \label{Matsubara3}  \\
\beta^{-1}\sum_n  \frac{1}{i\omega_n+\xi}\frac{1}{(i\omega_n-\xi)^3}  &=&-\frac{\tanh(\beta\xi/2)}{8\xi^3} -\frac{\beta}{16}\frac{d}{d\xi}\left(\frac{\mathrm{sech}^2(\beta\xi/2)}{\xi}\right).
\label{Matsubara4} 
\eqae

We make use of the integral
\eq \label{tanhInt}
\int_0^a \frac{X}{x} dx = \tanh(a)\ln(a)+\ln(4e^\gamma/\pi)+\int_a^\infty Y\ln(x) dx,
\eqe
where $X\equiv\tanh(x)$ and $Y\equiv \mathrm{sech}^2(x)$ and 
 $\gamma=0.577216\dots$ is the Euler-Mascheroni constant. Setting $a$ to 0 gives the result
\eq \label{sech2Int}
\int_0^\infty Y\ln(x)dx = -\ln(4e^\gamma/\pi).
\eqe
Also, the definite integral
\eq \label{XYx}
\int_0^\infty\frac{XY}{x} dx = \frac{7}{\pi^2}\zeta(3),
\eqe
can be combined with
\eq  \label{App_dDx_x}
\int_0^a \frac{dx}{x} \frac{d}{dx}\left(\frac{X}{x} \right) = 
\tfrac{1}{2}a^{-2}[\tanh(a)-a \ \mathrm{sech}^2(a)] - \int_0^a\frac{XY}{x} dx,
\eqe
to give
\eq \label{dtanhInt}
\int_0^\infty \frac{dx}{x} \frac{d}{dx}\left(\frac{ X}{x} \right) = - \frac{7}{\pi^2}\zeta(3).
\eqe
Using this
\eq\ \label{XxY}
\int_0^\infty x^{-3}[X-xY] dx = -\int_0^\infty \frac{dx}{x} \frac{d}{dx}\left(\frac{ X}{x} \right) = \frac{7}{\pi^2}\zeta(3).
\eqe
Principal value integrals with singularity and lower boundary occurring at $0$ can be treated with the
formula \cite{Ramm:1990fk}
\eq
\ddashint_0^a \frac{f(x)}{x^{p+1}} dx = \int_0^a \frac{1}{x^{p+1}}\left[ 
f(x) - \sum_{j=0}^p \frac{f^{(j)}(0)x^j}{j} \right]dx
+
\sum_{j=0}^{p-1}  \frac{f^{(j)}(0)}{j!}\frac{a^{-p+j}}{j-p}
+
\frac{f^{(p)}(0)}{p!}\ln(a) 
\eqe 
Taking $p=1$ and $p=2$ and $f(x)=X(x)$ gives

\begin{subequations}\label{Xx32Cauchy}
\begin{align}
\ddashint_0^a \frac{X}{x^{3}} dx &= 
-\tfrac{1}{2}a^{-2}[\tanh(a)+a\ \sech^2(a)] - \int_0^a \frac{XY}{x} dx, \\
\ddashint_0^a \frac{X}{x^{2}} dx &= 
1+\sech^2(a)\ln(a) - \frac{\tanh(a)}{a}+2\int_0^a XY\ln(x) dx.
\label{Hypersingx2Gen}
\end{align}
\end{subequations}
Using (\ref{Hypersingx2Gen}) we obtain
\begin{subequations}\label{Hypersingx2}
\begin{align}
\ddashint_0^\infty \frac{X}{x^{2}} dx & = 1 + 2\int_0^\infty XY\ln(x) dx \equiv \eta  ,
\label{Hypersingx2a}\\
\ddashint_0^a \frac{X}{x^{2}} dx &\approx \ln(a), \qquad\qquad (a\ll 1)
\label{Hypersingx2b} \\
\ddashint_0^\infty \frac{X}{x^{3}} dx & = -\frac{7\zeta(3)}{\pi^2} ,
\label{Hypersingx2c}\\
\ddashint_0^a \frac{X}{x^{2}} dx &\approx -a^{-1} . \qquad\qquad (a\ll 1) 
\label{Hypersingx2d} 
\end{align}
\end{subequations}
Here, $\eta\approx 0.7905$ is a constant.
For $a\gtrsim1$ one can use the leading term of the asymptotic series approximation
for the exponential integral, $E_1(z) = \frac{e^{-z}}{z}\sum_{n=0}^{N-1}\frac{n!}{(-z)^n}$,
with $N=1$ to derive
 the following useful approximations
\begin{align}
\int_a^\infty \frac{XY}{x} dx &= -4\sum_{n=1}^\infty (-1)^n n^2 E_1(2na)
\simeq \frac{1}{2a}\sech^2(a),
&
(a\gtrsim 1),
\label{XYxApprox} \\
a\int_a^\infty \frac{X}{x^2} dx  &= 1 - 2\sum_{n=1}^\infty (-1)^n \left[ e^{-2na}+2naE_1(2na) \right]
& \nn\\
&  \simeq 1+\frac{4}{1+e^{2a}} \approx 1. &
(a\gtrsim 1).
\label{Xx2AsymptoticSeries}
\end{align}


\section{Fierz identities}
\label{Fierz_Sec}

To derive the Fierz identities for the four-component representation, one can consider the basis:
\[
\lbrace I,\ \gamma^\mu,\ \gamma^{35}\gamma^\mu,\ \gamma^3\gamma^\mu,\ \gamma^5\gamma^\mu,\ i\gamma^3,\ i\gamma^5, \gamma^{35}  \rbrace.
\]
However, the bilinears involving $\gamma^3$ and $\gamma^5$ do not transform simply under parity and charge conjugation, but instead transform
as doublets. We have
\eq\label{axipseudoscalarDef}
\Phi = \begin{pmatrix}\bar\psi i\gamma^3\psi \\ \bar\psi i\gamma^5\psi  \end{pmatrix},
\qquad
\Phi^\mu = \begin{pmatrix}\bar\psi \gamma^3\gamma^\mu\psi \\ \bar\psi \gamma^5\gamma^\mu  \psi \end{pmatrix}, 
\eqe
which are termed axi-scalar and axi-vector respectively in \cite{Burden:1991mb}, as under parity they transform as
$(\Phi)^\mathcal{P} = R_P\Phi$ and $(\Phi^\mu)^\mathcal{P} = \Lambda^\mu{}_\nu R_P\Phi^\mu$ 
with $ \Lambda^\mu{}_\nu = \mathrm{diag}(1,-1,1)$ and $R_P
 = \left(\begin{smallmatrix}
 -\cos(2\phi_P)  & \sin(2\phi_P)  \\
  \sin(2\phi_P) & \cos(2\phi_P)
\end{smallmatrix} \right)$.
Introducing the shorthand
\[
\begin{array}{lll}
s_1 = (\bar\psi_4\psi_2)(\bar\psi_3\psi_1) & \qquad & s_2 = (\bar\psi_4\psi_1)(\bar\psi_3\psi_2) \\
v_1 = (\bar\psi_4 \gamma^\mu \psi_2)(\bar\psi_3  \gamma_\mu \psi_1) & \qquad & v_2 = (\bar\psi_4 \gamma^\mu \psi_1)(\bar\psi_3  \gamma_\mu \psi_2) \\
a_1 = (\bar\psi_4 \gamma^{35}\gamma^\mu \psi_2)(\bar\psi_3  \gamma^{35} \gamma_\mu \psi_1) & \qquad & a_2 = (\bar\psi_4  \gamma^{35} \gamma^\mu \psi_1)(\bar\psi_3  \gamma^{35} \gamma_\mu \psi_2) \\
x_1 = (\Phi^\dagger)_{(4;2)}(\Phi)_{(3;1)} & \qquad& x_2 = (\Phi^\dagger)_{(4;1)}(\Phi)_{(3;2)} \\
y_1 = (\Phi^\dagger_\mu)_{(4;2)}(\Phi^\mu)_{(3;1)} & \qquad& y_2 = (\Phi^\dagger_\mu)_{(4;1)}(\Phi^\mu)_{(3;2)} \\
z_1 = (\bar\psi_4 \gamma^{35} \psi_2)(\bar\psi_3 \gamma^{35} \psi_1) & \qquad & z_2 = (\bar\psi_4 \gamma^{35} \psi_1)(\bar\psi_3 \gamma^{35} \psi_2) 
\end{array}
\]
where the subscripts on the $\Phi$ and $\Phi^\mu$ terms indicate the arrangement of the fermion bilinears that comprise the doublet, the Fierz identities are\footnote{
The Fierz identities for the other two linear combinations of the $i\gamma^3$ and $i\gamma^5$, and 
$\gamma^{3}\gamma^\mu$ and $\gamma^{5}\gamma^\mu$ bilinears form a closed system
and need not be considered here.
}
\eq
\begin{pmatrix}
s_1 \\ v_1 \\ a_1 \\ x_1 \\ y_1 \\ z_1 
\end{pmatrix}
= -\frac{1}{4}
\begin{pmatrix}
\nomin1   & \nomin1  & \nomin1  & -1 & -1 & \nomin1 \\
\nomin3   & -1 & -1 & -1 &  \nomin3 & \nomin3 \\
\nomin3   & -1 & -1 &  \nomin1 & -3 & \nomin3 \\
-6  & -2 & \nomin 2 &  \nomin0 & \nomin0  & \nomin6 \\
-2  & \nomin2  & -2 &  \nomin0 & \nomin0  & \nomin2 \\
\nomin1   & \nomin1  & \nomin 1  & \nomin 1 & \nomin 1 &\nomin 1 \\
\end{pmatrix}
\begin{pmatrix}
s_2 \\ v_2 \\ a_2 \\ x_2 \\ y_2 \\ z_2 
\end{pmatrix}.
\label{FierzMatrix}
\eqe
When the four fermion fields are equal $\psi_1 = \psi_2 = \psi_3 = \psi_4 =\psi $, we can drop the subscripts on the variables $s_1$ to $z_1$ and obtain the relationship
\eq
3(s+z)+(a+v)=0. \label{noindexFierz}
\eqe
Since the bilinears $s$ and $z$ are even under charge conjugation, and $v$ and $a$ are odd, we can conjugate one of the bilinears in each combination to get
\eq
3(s_1+z_1)-(a_1+v_1)=0 ,
\eqe
where $\psi_2 = \psi_4=\psi$ and  $\psi_3 = \psi_1=\psi^c$.
Using (\ref{FierzMatrix}) we find
\eq
v_2 + a_2 = 0,
\eqe
which can be substituted back into the identity for $z_1$ to give
\[
m = m_1 = -\tfrac{1}{4}[s_2 + x_2 + y_2 +z_2].
\]
More explicitly, this is 
\eq
(\bar\psi \gamma^{35}\psi)^2 = -\tfrac{1}{4}\left[ (\bar\psi \psi^c)(\bar\psi^c \psi)
+  (\bar\psi \gamma^{35} \psi^c)(\bar\psi^c   \gamma^{35} \psi)
+ \tilde\Phi^\dagger \tilde\Phi +  \tilde\Phi^\dagger_\mu \tilde\Phi^\mu 
  \right],
  \label{zExpansion}
\eqe
with
\eq
\tilde\Phi = \begin{pmatrix}\bar\psi^c i\gamma^3\psi \\ \bar\psi^c i\gamma^5\psi  \end{pmatrix},
\qquad
\tilde\Phi^\mu = \begin{pmatrix}\bar\psi^c \gamma^3\gamma^\mu\psi \\ \bar\psi^c \gamma^5\gamma^\mu  \psi \end{pmatrix}.
\label{tildePhiDef}
\eqe

\section{Expansion terms}
\label{KinExpansion}

The quadratic part of the spatial Lagrangian can be
expanded to second order in gradients as
\[
\left[ 
 M_{Q_1}\nabla^2 + M_{Q_2} (\vec k \cdot \vec\nabla)^2  
-\frac{1}{\omega_n^2+\xi^2} -\frac{1}{\omega_n^2+\bar\xi^2} \right]|\Delta_A|^2,
\]
where
\begin{subequations}\label{RelKineticTerms}
\begin{align}
M_{Q_1}& = -\frac{(\omega_n^2+\xi\bar\xi)(2\omega_n^2+\xi^2+\bar\xi^2)}{
(\omega_n^2+\xi^2)^2(\omega_n^2+\bar\xi^2)^2}, \\
M_{Q_2}& =
 \frac{4[(\omega_n^2+\xi\bar\xi)^2 - \omega^2_n(\xi-\bar\xi)^2]
(2\omega_n^2+\xi^2+\bar\xi^2)}{
(\omega_n^2+\xi^2)^3(\omega_n^2+\bar\xi^2)^3} -
\frac{4(\omega_n^2+\xi\bar\xi)}{(\omega_n^2+\xi^2)^2(\omega_n^2+\bar\xi^2)^2}.
\end{align}
\end{subequations}
The relevant Matsubara sums are
\begin{subequations}\label{KineticTermMatSums}
\begin{align}
\beta^{-1}\sum_n M_{Q_1} & =
\frac{\beta}{8(\xi+\bar\xi)}\left[
\frac{\sech^2(\tfrac{1}{2}\beta\xi)}{\xi} + \frac{\sech^2(\tfrac{1}{2}\beta\bar\xi)}{\bar\xi} \right] 
- \nn\\
&  \quad
-\frac{1}{4(\bar\xi-\xi)}\left[
\frac{\tanh(\tfrac{1}{2}\beta\xi)}{\xi^2} - \frac{\tanh(\tfrac{1}{2}\beta\bar\xi)}{\bar\xi^2} \right], \\
\beta^{-1}\sum_n M_{Q_2} & =
-\frac{\beta^2}{4(\xi+\bar\xi)^2}\left[
\frac{\sech^2(\tfrac{1}{2}\beta\xi)\tanh(\tfrac{1}{2}\beta\xi)}{\xi}
 + \frac{\sech^2(\tfrac{1}{2}\beta\bar\xi)\tanh(\tfrac{1}{2}\beta\bar\xi)}{\bar\xi} 
\right] -
\nn\\
& \quad
-\frac{\beta}{4(\xi+\bar\xi)^3}\left[
\frac{\bar\xi+3\xi}{\xi^2}\sech^2(\tfrac{1}{2}\beta\xi) + \frac{3\bar\xi+\xi}{\bar\xi^2}\sech^2(\tfrac{1}{2}\beta\bar\xi) 
\right] + \nn\\
& \quad
+\frac{1}{2(\bar\xi^2-\xi^2)}\left[
\frac{\tanh(\tfrac{1}{2}\beta\xi)}{\xi^3} - \frac{\tanh(\tfrac{1}{2}\beta\bar\xi)}{\bar\xi^3} \right].
\end{align}
\end{subequations}

\section{The time-dependent Lagrangian to second order }
\label{TimeDerivativeSecondOrder}

\subsection{Nonrelativistic case}

Using (\ref{Hypersingx2}), the terms second order in $q_0$ in (\ref{nonRel_QDefs})
are\footnote{
There is no $Q''(q_0)$ term as the next to leading order term in the
 expansion of the $\tanh$ function in (\ref{NonRelTimeTanhTerm}) in $\mathcal{O}(q_0^3)$.
}
\eq
\begin{split}
Q'(q_0)  & \supset \frac{\nu(0)}{2} 
\frac{\alpha^2\beta^2 q_0^2}{16} \left[
\frac{1}{\alpha^4}
\ddashint_{-c_1}^0 \frac{X}{x^3} dx  +
\ddashint_0^\infty \frac{X}{x^3} dx
\right] ,
 \\
& \simeq
-\frac{\nu(0)}{2}
 \frac{\alpha^2\beta^2 q_0^2}{16} \left[ \frac{7\zeta(3)}{\pi^2}+\frac{1}{\alpha^4c_1}
\right] ,  \\
& \simeq
-\frac{\nu(0)}{2} 
 \frac{\alpha^2\beta^2 q_0^2}{16},
\end{split}
\eqe
where 
in the third line, we have neglected the subdominant term
 $1/(\alpha^{4}c_1) = 1/(\alpha^{2}c_2) \ll 1 $ as $\alpha$ is large.
After integrating by parts, the dynamical part of the Lagrangian to second order is 
\eq\label{Leff2_Time_NonRel}
\mathcal{L}^{(2)}_{\rm dyn} \simeq \left(\frac{m}{2\pi}\right)
\frac{7\alpha^2\beta^2\zeta(3) }{32\pi^2}  
\Delta^*\p_0^2 \Delta .
\eqe

\subsection{Relativistic case}

If we were to expand up to second order in $q_0$ in (\ref{Q''+-Terms}) the integration by parts gives a relative plus
sign. The $q_0^2$ terms cancel exactly and we have only to evaluate the hypersingular integral $Q_\xi'(q_0)$ and the integral $Q_{\bar\xi}'(q_0)$, given by
\eq
\begin{split}
Q_\xi'(q_0) &= \frac{q_0^2 \alpha\beta}{32\pi} \left\lbrace \frac{1}{\alpha^2}\ddashint_{-c_1}^0 dx \frac{(x+c_1)X}{x^3} + \ddashint_{0}^\infty dx \frac{(x+c_2)X}{x^3}\right\rbrace, \\
& \simeq 
 -\frac{q_0^2 \alpha\beta}{32\pi} \left\lbrace
 \frac{1}{\alpha^2} [1+\ln(c_1)]-\eta + \frac{7\zeta(3)}{\pi^2}c_2 
\right\rbrace,
\end{split}
\label{QDashxi}
\eqe
where in the second line we have evaluated the principal value
integrals using (\ref{Xx32Cauchy}), keeping only the non-vanishing terms
as $c_1\rightarrow 0$, and 
\eq
\begin{split}
Q_{\bar\xi}'(q_0) &=
\frac{q_0^2 \alpha\beta}{32\pi} \left\lbrace \frac{1}{\alpha^2}\ddashint_{c_1}^{2c_1} dx \frac{(x-c_1)X}{x^3} + \ddashint_{2c_2}^\infty dx \frac{(x-c_2)X}{x^3}\right\rbrace, \\
& \simeq
\frac{q_0^2 \alpha\beta}{32\pi}\left\lbrace \frac{1}{\alpha^2}
\left[ \ln 2 -\frac{1}{2} \right]
+\frac{3}{8c_2}\tanh(2c_2)-\sech^2(2c_2)\ln(2c_2)+2\int_{2c_2}^\infty XY\ln(x) dx
\right\rbrace, \\
& \lesssim
\frac{q_0^2 \alpha\beta}{32\pi}\left\lbrace \frac{1}{\alpha^2}
\left[ \ln 2 -\frac{1}{2} \right] + \frac{1}{2} \right\rbrace.
\end{split}
\label{QDashbarxi}
\eqe
This is valid for $c_2\gtrsim 1$ as in the second line we have integrated by parts and used (\ref{XYxApprox}), and in the third line noted that the expression is a decreasing function of $c_2$. 
Combining (\ref{QDashxi}) and (\ref{QDashbarxi}),
we have
\eq
Q_{\xi}'(q_0) + Q_{\bar\xi}'(q_0) \approx 
 -\frac{q_0^2 \alpha\beta}{32\pi} \left(
 \frac{\ln(c_1)}{\alpha^2}+ \frac{7\zeta(3)}{\pi^2}c_2 
\right)
\approx
 -\frac{q_0^2 \alpha\beta}{32\pi}\frac{7\zeta(3)}{\pi^2}c_2,
\eqe
modulo corrections by subdominant numerical factors $\sim\mathcal{O}(1)$ inside the parentheses.
In the second approximation, we have used the fact that for large $\alpha$ we have
\[
 \frac{\ln(c_1)}{\alpha^2}+ \frac{7\zeta(3)}{\pi^2}c_2  =   \frac{\ln(c_2/\alpha^2)}{\alpha^2}+ \frac{7\zeta(3)}{\pi^2}c_2 \rightarrow \frac{7\zeta(3)}{\pi^2}c_2,
\]  
so the second term is dominant.
The correction to the Lagrangian (\ref{Rel_LDyn2}) is then
\eq
\mathcal{L}_{\rm dyn}^{(2)} \approx  
\left(\frac{\mu}{2\pi}\right)
\frac{7\alpha^2\beta^2\zeta(3) }{32\pi^2}  
 \Delta_A^\dagger
 \p_0^2   \Delta_A.
\eqe
Comparing with (\ref{Leff2_Time_NonRel}) it can be seen that the scale $\mu$
enters in the same way as the mass of the fermions in the nonrelativistic case. 
In the nonrelativistic case, the $\mathcal{L}_{\rm dyn}^{(2)}$ represents the second
term in series expansion in $q_0/T$, which appears in the Lagrangian as the Cooper pairs can
break up. The presence of an equivalent term in the relativistic case suggests that the weakly interacting bosons described by (\ref{RelCase_Final_L}) are not stable, although this is most
important for the higher energy modes. 

\bibliographystyle{h-physrev}
\bibliography{BVRefs}

\end{document}